\documentclass[11pt]{article}

\usepackage{amsmath}
\usepackage{amssymb}
\usepackage{amsfonts}
\usepackage{mathrsfs}
\usepackage{latexsym}
\usepackage{graphicx}
\usepackage{amsthm}
\usepackage{caption}
\usepackage[font=small,labelfont=bf]{caption}
\usepackage{xcolor}
\usepackage{subcaption}
\usepackage[text={16cm,25cm}]{geometry}
\usepackage{mathptmx}

\newcommand\ignore[1]{}

\renewcommand{\epsilon}{\varepsilon}
\numberwithin{equation}{section}

{\begingroup

  \begin{enumerate}}
  {\end{enumerate}\endgroup}

\title{Analysis and Simulation of Extremes and Rare Events in Complex Systems.}
\author{Meagan Carney\thanks{Max Planck Institute for the Physics of Complex Systems,
N\"othnitzer Str.\ 38, D 01187 Dresden, Germany. Email: meagan@pks.mpg.de } \and
Holger Kantz\thanks{Max Planck Institute for the Physics of Complex Systems,
N\"othnitzer Str.\ 38, D 01187 Dresden, Germany. Email: kantz@pks.mpg.de }
\and Matthew Nicol\thanks{Department of Mathematics, University of Houston, Houston TX 77204-3008, USA.  Email: nicol@math.uh.edu}}
\date{\today}
\begin{document}
\maketitle

\begin{abstract}

Rare weather and climate events, such as heat waves and floods, can  bring tremendous social costs. Climate data is often limited in duration 
and spatial coverage, and climate forecasting has often  turned to simulations of climate models to make better predictions of rare weather events. However
  very long simulations of complex models, in order  to obtain accurate probability estimates, may be prohibitively slow. It is an important scientific
problem to develop probabilistic and dynamical techniques to estimate the probabilities of rare events accurately from limited data. 
In this paper we compare four modern methods of estimating the probability of rare events: the generalized extreme value (GEV) method from classical extreme value theory;  two importance sampling techniques,  geneaological particle analysis (GPA) and the Giardina-Kurchan-Lecomte-Tailleur (GKLT) algorithm; as well as brute force Monte Carlo (MC). With these techniques we  estimate the probabilities of rare events in three dynamical models: the Ornstein-Uhlenbeck process, the Lorenz '96 system and PlaSim (a climate model).  We keep the computational effort constant and see how well the rare event probability estimation of each technique compares to a gold standard afforded by a very long run control. Somewhat 
  surprisingly we find that classical extreme value theory  methods outperform GPA,  GKLT and MC at estimating rare events.

\end{abstract}


\section{Extremes and rare event computation.}

  Rare weather and climate events such as   heat waves,  floods,  hurricanes, and the like,   have  enormous social and financial consequences. It is important to be able to estimate as accurately
  as possible the probability of the  occurrence and duration of such extreme events.
  However  the time series data available  to predict rare events is  usually too short to assess with reasonable  confidence the probability of  events with very long recurrence times, for example
  on the order of  decades or centuries. In this regard, one may consider return levels of exceedances which represent the level that is expected to be exceeded on average once every 100 years by a process. For example, a 100-year return level estimate of a time series of temperature or precipitation data would tell us the temperature or amount of precipitation that is expected to be observed only once in 100 years. It is common, however, that the amount of weather data available is limited in spatial density and time range. As a result, climate forecasting has often turned to simulations of climate models to make better predictions of rare weather events. These simulations are not without limitations; a more accurate model requires a large amount of inputs to take into account most of the environmental factors which impact weather. With these more complex models, very long simulations may be required to obtain probability estimates of rare events with  long return times. These simulations may be very slow and have motivated the study of statistical techniques which allow for more accurate rare event probability estimates with lower computational cost.
 
 One approach to estimate the probability of rare events or extremes is to use classical extreme value theory, perhaps aided by clustering techniques or other statistical approaches suitable for the application at hand. 
Other techniques to accurately estimate the probabilities of rare events include importance sampling (IS) methods. In general, importance sampling is a probabilistic technique which allows us to choose those trajectories or paths in a random or deterministic model which will most likely end in an extreme event. This reduces the number of long trajectories that are required to obtain an estimate on the tail probabilities of extremes and essentially changes the sampling distribution to  make rare events less rare.  The goal of  importance sampling is not only to estimate probabilities of rare events with less 
computational cost, but also more accurately in that the ratio of the likely error in estimation   to the probability of the event is lessened.

  Importance sampling algorithms have been successfully applied in many fields, especially in chemical and statistical physics~\cite{rubino_tuffin,del_moral1,bucklew}. Recently 
  these techniques have been applied to dynamical systems and dynamical climate models~\cite{Bouchet1, Bouchet2}. In this paper we will consider two similar types
  of IS techniques,  geneaological particle analysis (GPA) and the Giardina-Kurchan-Lecomte-Tailleur (GKLT) algorithm. The GKLT algorithm is designed  to estimate probabilities of events 
  such as heatwaves as it considers time-averaged quantities. GKLT is motivated by ideas from large deviations theory, though in its implementation it does not explicitly require calculation of large
  deviation quantities such as rate functions.

  The main goal of this paper is to compare the performance of the generalized extreme value (GEV) method with GPA,  GKLT and brute force Monte Carlo (MC) at estimating rare events of our test models: the Ornstein-Uhlenbeck process, the Lorenz '96 system and PlaSim (a climate model).
  We keep the computational effort constant and see how well the rare event probability estimation of each technique compares to a gold standard afforded by a very long run control. Somewhat surprisingly we find that GEV outperforms GPA,  GKLT and MC at estimating rare events. Perhaps this advantage comes from the fact that GEV methods are parametric and maximum likelihood estimation,  in practice, results in close to optimal parameters and confidence intervals.

 
\section{The Four Methods.}

Extreme value theory  is a well-established branch of statistics~\cite{Gumbel,Leadbetter,Coles}. Over the last ten years or so the theory has been investigated in
the setting of chaotic dynamics, for a state of the art review see~\cite[Chapters 4 and 6]{Book}. The goal of extreme value theory is to estimate probabilities associated to rare events. Another way to approach this problem is via  importance sampling. Recently  ideas from importance sampling have been successfully applied to several dynamical models (a non-exhaustive list includes~\cite{Bouchet_Wouters,Garnier_DelMoral,GKP, KT}).  How do the methods compare, for a given computational cost, at accurately determining the
probabilities of rare events? We now describe the four methods we investigate in this paper.

  \subsection{Generalized Extreme Value Distribution (GEV).}
  There are two main approaches
  for classical extreme value theory: peaks over threshold; and the block maxima method. 
 They are equivalent mathematically~\cite{Coles}, but more research
  has been done on the block maxima method in the setting of deterministic models (for a treatment of this topic and further references see~\cite[Chapters 4 and 6]{Book}). We will use the block maxima method in this paper. In the context of modeling extremes in dynamical models, 
  Galfi et al~\cite{Galfi} have used the peaks over threshold method to benchmark their 
  large deviations based analysis of heat-waves and cold spells in the PUMA model of atmospheric circulation. 
  Given a sequence of iid random variables $\{X_1,X_2, \ldots, X_n, \ldots\}$ it is known that the maxima process $M_n=\max\{X_1,X_2,\ldots,X_n\}$ has only  three possible non-degenerate limit distributions under linear scaling: Types I (Gumbel), II (Fr\'echet) and III (Weibull)~\cite{Galambos}, no matter the distribution of $X_1$. By linear scaling we mean the 
  choice of a sequence of constants $A_n$, $B_n$ such that $P(A_n (M_n-B_n) \le y)\rightarrow H(y)$ for a nondegenerate distribution $H$. The extreme value distributions 
  are universal and play a similar role to that of the Gaussian distribution in explaining a wide variety of phenomena. These three distributions can be subsumed into a Generalized Extreme Value (GEV) distribution
 \[
 G(x) =\mbox{exp}\Big( -[1+\zeta\big(\frac{x-\mu}{\sigma}\big) ]^{\frac{-1}{\zeta}}\Big)~(*)
 \]
 defined for $\{ x : 1+\zeta\big(\frac{x-\mu}{\sigma}\big) >0\}$ with three parameters $-\infty < \mu < \infty$, $\sigma>0$, $-\infty < \zeta < \infty$. The parameter $\mu$ is the location parameter, $\sigma$ the scale and $\zeta$ the shape parameter (the most important parameter as $\zeta$ determines the tail behavior).  A type I distribution corresponds to the limit as $\zeta \to 0$, while Type II corresponds to $\zeta>0$ and Type III to  $\zeta <0$. The three types differ in the behavior of the tail of the distribution function $F$ for the underlying process $(X_i)$. For type III the $X_i$ are essentially bounded, while the tail of $F$ decays exponentially for Type I and polynomially (fat tails) for Type II. 
 
 The advantage of using GEV over brute force fitting a tail distribution by simulation or data collection is that a statistical distribution is assumed, and only three parameters need to be determined (like fitting a normal distribution, where only 2 parameters need to be estimated). This has enormous advantages over methods which try to determine an a priori unknown form of distribution.  The GEV parameters may be estimated, for example, 
 by the method of maximum likelihood. 
 Once the parameters are known $G(x)$ can be used to make predictions about extremes.  This is done for a time series of  observations in the following way. A sequence of observations are taken $X_1$, $X_2$, ... and grouped into blocks of length $m$ (for example it could be daily rainfall amounts clumped into blocks of one year length). This gives  a series of block maxima $M_{m,1}$, $M_{m,2}$, ... where $M_{m,\ell}$ is the maximum of the observations in block $\ell$ (which consists of $m$ observations). Using parameter estimation like maximum likelihood, the GEV model is fitted to the sequence of $M_{m,\ell}$ to yield $\mu$, $\sigma$ and $\zeta$. The probability of certain return levels of exceedance for the maximum of time-series of length $m$ are obtained by inverting $(*)$ and subtracting from $1$. For example, $m$ could correspond to a length of one year made of $m=365$ daily rainfall data points, then the result is the level of rainfall $a$ that the yearly maximum is expected to exceed once every $1/(1-G(a))$ years.

One issue in the implementation of GEV  is the possibly slow rate of convergence to the limiting distribution. There are some results~\cite{Holland_rates, FFT_rates} on rates of convergence
to an extreme distribution for chaotic systems, but even in the the iid case rates of convergence may be slow~\cite{Hall}.   Another is the assumption of independence. Time-series from weather readings, climate models or deterministic dynamical systems 
are usually highly correlated. There are conditions in the statistical literature~\cite{Leadbetter, Collet, FFT, GHN} under which the GEV   distributional limit holds for maxima $M_n$ of  observables $\phi(X_j)$ which are ``weakly dependent''  i.e. the underlying  $X_j$ are correlated, and which ensure that $M_n$ has the same extreme value limit law as an iid process with the same distribution function. Usually 
two conditions are given, called  Condition $D_2$  (a  decay of correlations requirement), and Condition $D^{'} $ (which quantifies short returns) which need to be checked. Collet~\cite{Collet} first used
variants of Condition $D_2$ and Condition $D^{'} $  to establish return time statistics and extremes for certain dynamical systems. Recent results~\cite{Book} have shown that maxima of time-series of H\"older observables on a wide variety of chaotic dynamical systems (Lorenz models, chaotic billiard systems, logistic-type maps and other classical systems) satisfy classical extreme value laws.
The development of extreme value theory for deterministic dynamical systems has been an intensive area of research. For the current state of knowledge we refer to ``Extremes and Recurrence in Dynamical Systems''~\cite[Chapters 4 and 6]{Book}. 

 Even  using a parametric model like GEV  there is still an issue of having enough data. There are several approaches to extract the most information possible from given measurements.  For example,  in~\cite{CAN, Carney_Kantz} sophisticated clustering techniques based on information theory ideas were used to group measurements
  from different spatial  locations  and amplify time-series of temperature recordings to
  improve the validity of GEV estimates for annual summer temperature extremes in Texas and Germany.
  
  Despite these caveats this paper shows that GEV works very well in estimating probabilities of rare events in realistic models such as PlaSim, performing better at the same computational cost than MC and  the 
  two IS techniques we investigate.
  

  \subsection{Brute Force Monte Carlo.}
  
  Given a random variable $X$ distributed according to a distribution $\rho(x)$, we want to estimate the probability of a rare event,
  \[
  \gamma_A = P(X\in A)<<1
  \]
  As a naive approach, one could do this by a brute force Monte Carlo (MC) estimate,
  \[
  \hat{\gamma_A} (N) = \frac{1}{N}\sum_{i=1}^N 1_{A}(X_i) 
  \]
  for some sequence of random variables $X_i$ sampled from $\rho(x)$. Here, $E(\hat{\gamma_A})=\gamma_A$ (as $\hat{\gamma_A}$ is an unbiased estimator) and for large enough $N$,
  \[
  \sqrt{N}\hat{\gamma_A}(N) \sim \mathcal{N}(\gamma_A,\sigma^2(\hat{\gamma_A}))
  \]
  by the central limit theorem (where valid).  The relative error of an estimator is defined to be the standard deviation of the estimator divided by the 
  estimated quantity. As
  \[
  \sigma^2(1_A) = E((1_A(X)-\gamma_A)^2)=E(1_A(X))-\gamma^2_A = \gamma_A-\gamma^2_A\approx \gamma_A
  \]
  for small $\gamma_A$, and  $\mbox{Var}\hat{\gamma_A} (N)=\frac{\mbox{Var}\gamma_A}{N}$,
  the relative error is estimated as 
  \[
  \sigma(\hat{\gamma_A})(N) /\gamma_A\approx\frac{1}{\sqrt{N\gamma_A}},
  \]
  which is large for  small $\gamma_A$.   This analysis can be found in \cite{Bouchet_Wouters}. 
  
\subsection{Importance Sampling Techniques}

Importance sampling methods work to lower the relative error by a change of measure from $\rho$ to another measure $\tilde{\rho}$. 
The idea is to change the distribution of $X$ in a way that rare events are sampled more often under $\tilde{\rho}$
and if the steps in the algorithm by which we do this are accounted for, we obtain an accurate estimate of  the probability of the rare event under $\rho$ with a
significantly decreased relative error in our estimator.  In our applications $X$ is a real-valued random variable (distance from the origin in Ornstein-Uhlenbeck process, energy level in the Lorenz '96 model and temperature or averaged temperature in the PlaSim model) and rare events will correspond
to high values of $X$.

We alter the probability of rare events  by using a weight function whose goal is to perform a change of measure. Provided $X$ has tails which decay exponentially, the weight function can be chosen as an ``exponential tilt''. We now provide an illustration of the exponential tilt in the context of a normally distributed random variable. Details for the following  estimates are provided in~\cite{Bouchet_Wouters}.

Suppose we want to estimate the probability $\gamma_A$ of a rare event $A = \{X>a\}$ for $X\sim \mathcal{N}(0,1)$ so that $\rho(X)=e^{-x^2/2}$. If we choose,
\begin{equation}\label{eq.1}
\tilde{\rho}(X)=\frac{\rho(X)e^{CX}}{E(e^{CX})} = \frac{1}{\sqrt{2\pi}}\exp[\frac{-(X-C)^2}{2}]
\end{equation}
we obtain a shift of the average by $C$. The error of our estimate in the shifted distribution is given by its variance,
\[
\sigma^2(\tilde{\gamma}_A) = P_{C,1}(X>a)e^{C^2}-\gamma^2_A
\]
where $P_{C,1}$ denotes the probability under a normal distribution with mean $C$ and variance $1$. 
If we take $a=2$ there is a unique minimum of the variance for a value of $C$ close to $2$. In this example a decrease of relative error by a factor of roughly $4$ is 
produced. Because of the scaling $\frac{1}{\sqrt{N\gamma_A}}$ it would take a 16 times longer brute force run to achieve this result. We remark that this exponential tilt of the original distribution results in an optimal value of $C(a)$ for each threshold $a$ for which $\gamma_A = P(X>a)$.  Part of the finesse in using IS techniques is to tune the parameter $C$. 

We now describe the two importance sampling techniques we investigate.

\subsubsection{Genealogical Particle Analysis}

Genealogical particle analysis (GPA)~\cite{Bouchet_Wouters,Garnier_DelMoral} is an importance sampling algorithm that uses weights to perform a change of measure, by a weight function $V(x)$ (in the previous example $V(x)$ was taken to be $Cx$ but V(x)  is application specific)  the original distribution of particles $x_t$ under the dynamics. When we talk of particles we may mean 
paths in a Markov chain model or trajectories in a dynamical model such as the Ornstein-Uhlenbeck process or Lorenz '96. These weights can be thought of as measuring the performance of a particle's trajectory. If the particle is behaving as though it comes from the distribution tilted by the weight function $V(x)$ then it is cloned, otherwise it is killed and no longer used in the simulation. The act of killing or cloning based on weights is performed at specified time steps separated by length $\tau$. We will refer to $\tau$ as the \textit{resampling time}. In theory, the resampling time can chosen between the limits of the Lyapunov time, so as to not be too large that samples relax back to their original distribution and the decorrelation time, so as to not be too small that all clones remain close to each other. In practice, the decorrelation rate of a trajectory $x_t$ under the dynamics is calculated as the autocorrelation taken over a time lag and the sampling time is then chosen as the smallest time lag that results in the autocorrelation of $x_t$ being
 close to  zero at a specified tolerance. A description of the algorithm is given below.


\begin{itemize}\label{gpa}
\item[1.] Initiate $n=1,\ldots, N$ particles with different initial conditions. 
\item[2.] For $i = 1,\dots,T_f/\tau$ where $T_f$ is the final integration time.
\begin{itemize}
	\item[2a.] Iterate each trajectory from time $t_{i-1}=(i-1)\tau$ to time $t_i=i\tau$.
	\item[2b.] At time $t_i$, stop the simulation and assign a weight to each trajectory $n$ given by,
	\begin{equation}
	W_{n,i} = \frac{\exp(V(x_{n,t_i})-V(x_{n,t_{i-1}}))}{Z_i}
	\end{equation}
	where
	\begin{equation}
	Z_i = \frac{1}{N}\sum_{n=1}^N W_{n,i}
	\end{equation}
	is the normalizing factor that ensures the number of particles in each iteration remains constant.
	\item[2c.] Determine the number of clones produced by each trajectory,
	\begin{equation}
	c_{n,i} = \lfloor W_{n,i} + u_n\rfloor
	\end{equation}
	where $\lfloor\cdot\rfloor$ is the integer portion and $u_n$ are random variables generated from a uniform distribution on $[0,1]$.
	\item[2d.] The number of trajectories present after each iteration is given by,
	\begin{equation}
	N_i = \sum_{n=1}^N c_{n,i}
	\end{equation}
	Clones are used as inputs into the next iteration of the algorithm. For large $N$, the normalizing factor ensures the number of particles $N_i$ remains constant; however, in practice the number of particles fluctuates slightly on each iteration $i$. To ensure $N_i$ remains constant it is common to compute the difference $\Delta N_i = N_i-N$. If $\Delta N_i>0$, then $\Delta N_i$ trajectories are randomly selected (without replacement) and killed. If $\Delta N_i<0$, then $\Delta N_i$ trajectories are randomly selected (with replacement) and cloned.
\end{itemize}
\item[3.] Provided $\tau$ is chosen between the two bounds specified above, the final set of particles tends to the new  distribution affected by $V(x)$ as $N\rightarrow\infty$,
\begin{equation}
\tilde{p}(x)=\frac{p(x)e^{V(x)}}{E(e^{V(x)})}.
\end{equation}
where $p(x)$ is the original distribution of the sequence of realizations $x_{n,0}$ and $\tilde{p}(x)$ is the distribution tilted by the weight function $V(x)$.
\end{itemize} 

\begin{figure}[h!]
	\centering
	\includegraphics[width=0.5\textwidth]{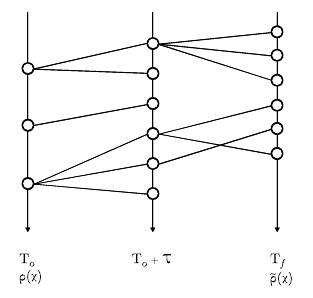}
	\caption{Illustration of the GPA algorithm. \label{fig:GPA}}
\end{figure}

Probability estimates for rare events $\gamma_A = P(X>a)$ under $p(x)$ are obtained by the reversibility of the algorithm and dividing out the product of weight factors applied to the particles. Suppose $A$ is the event $(X>a)$ for $X\sim p(x)$, then the expected value in the original distribution denoted by $E_0$ is given by~\cite{Bouchet_Wouters},
\begin{equation}
\gamma_A = E_0(1_{A}) = \frac{1}{N}\sum_{n=1}^N1_{A}(x_{n,T_f/\tau})e^{V(x_{n,0})}e^{-V(x_{n,T_f/\tau})}\prod_{i=1}^{T_f/\tau} Z_i
\end{equation}
Since GPA weights consider the end distribution of particles, they result in a telescoping sum in the exponential where the final rare event estimate is a function of the first and last weight terms only. For a detailed proof of this equivalence, we refer the reader to \cite{Bouchet_Wouters}. For an illustration of this algorithm, see fig. \ref{fig:GPA}.

As seen above, the change of measure is completely determined by the choice of weight function $V(x)$ in the algorithm. 


Furthermore, the algorithm can be applied to any observable $\phi$  by considering the  continuous random variable  $X_t = \phi(x_t)$ and defining 
\[
W_{n,i} = \frac{\exp(V(\phi(x_{n,t_i}))-V(\phi(x_{n,t_{i-1}}))}{Z_i}.
\]
where $x_n (t)$ is one of our $n=1,\ldots, N$ realizations and $x_{n,t_i}=x_n (t_i)$.

If we are interested in estimating rare event probabilities of a time-averaged quantity the weight $W_{n,i} = \frac{V(\int_{t_{i-1}}^{t_i}x_n(t) dt)}{Z_i}$ is given by an integral rather than the difference $W_{n,i} = \frac{\exp(V(x_{n,t_i})-V(x_{n,t_{i-1}}))}{Z_i}$ and  the 
increments do not telescope. We next discuss a method, the GKLT algorithm,  based on large deviations theory to estimate probabilities of rare events for time-averaged quantities in the next section. We note here that the GKLT algorithm in its implementation does not require explicit computation of large deviation quantities such as rate functions. 



\subsubsection{Giardina-Kurchan-Lecomte-Tailleur (GKLT) algorithm}

This technique was developed  in a series of papers~\cite{GKP, KT,GKLT} and uses ideas from large deviations theory to make estimates of extremes for time-averaged quantities, for example
heatwaves lasting a couple of months or more where the averaged maximal daily temperature over the two month period would be high. The advantage is that
over long periods of averaging  large deviation theory gives a method which works well, but a disadvantage is that the period of averaging needs to be long enough for the heuristic arguments  involving the  rate function and other quantities from large deviations theory to be valid. In practice, to calculate the probability of summer heatwave extremes in Europe,  the duration of heatwaves  has been set at  the order of 90 to 120 days in the literature~\cite{Galfi, Ragone_Wouters_Bouchet}. 


Suppose  $\phi$ is an observable. We will consider time-averaged quantities $\frac{1}{T} \int_{t=jT}^{(j+1)T}\phi(x(t))~dt$ over
a fixed time-window of length $T$, $j=1,\ldots, \lfloor T_f/T\rfloor$. We may choose to apply the weight function $V$ to the integral of $n = 1,\ldots, N$ realizations $\phi(x_n(t))$ by defining the set of weights as,

\begin{equation}\label{w:gklt}
W_{n,i} = \frac{V(\int_{t_{i-1}}^{t_i}\phi(x_n(t))~dt)}{Z_i}
\end{equation}
with normalizing factor,
\[
Z_i = \frac{1}{N}\sum_{n=1}^N W_{n,i}
\]
where the resampling time $\tau = t_{i-1}-t_i$ is chosen between the limits described in sec. \ref{gpa} and may differ from the choice of the time-average window length $T$.

\begin{figure}[h!]
	\centering
	\begin{minipage}{0.45\textwidth}
		\centering
	\includegraphics[width=\textwidth]{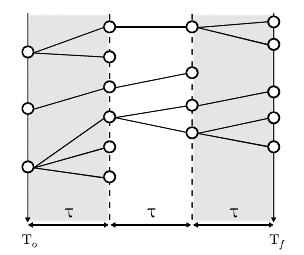}
	\caption*{(a)}
	\end{minipage}
	\begin{minipage}{0.45\textwidth}
	\centering
	\includegraphics[width=\textwidth]{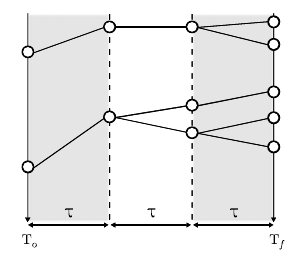}
	\caption*{(b)}
	\end{minipage}	
	\caption{(a) Illustration of the GKLT algorithm and (b) assembly of $N$ backward trajectories. Although shifts in the distribution of the integral are defined by the resampling time $\tau$, reconstruction of backward trajectories allows for estimates on $T$-time averaged trajectories after implementation of GKLT.\label{fig:GKLT}}
\end{figure}

Applying the method described in algorithm \ref{gpa} equipped with eq. \ref{w:gklt} tilts the distribution of the integral $\int_{t_{i-1}}^{t_i}\phi(x(t))~dt$ by $V(\cdot)$. As a result, the distribution of the $T$-time average trajectory $\frac{1}{T}\int_{t=jT}^{(j+1)T}\phi(x(t)) ~dt$ is tilted in a similar way. For an illustration of this algorithm, see fig. (\ref{fig:GKLT}).




Since the weight is a function of segments of the trajectory (rather than the distribution of end particles), the telescoping property no longer holds and estimates in the original distribution require the reconstruction of $N$-backward trajectories $\hat{\phi}(x_n(t))$, $n=1,\ldots, N$. 

 Let $E_0$ denote the expected value in the original distribution and suppose $O$ is some functional of  $\phi(x_n(t))$. Then it can be shown \cite{Ragone_Wouters_Bouchet},
\begin{equation}\label{eq.9}
E_0(O(\{\phi(x_n(t))\}_{0\le t\le T_f})) \sim \frac{1}{N}\sum_{n=1}^N O(\{\hat{\phi}(x_{n}(t))\}_{0\le t\le T_f}) e^{-V(\int_0^{T_f} \hat{\phi}(x_n(t))dt)}\prod_{i=1}^{T_f/\tau} Z_i.
\end{equation}

Often, $O$ in eq. \ref{eq.9} is taken as some indicator function of a rare event so that, $E_0(O(\{\phi(x(t))\}_{0\le t\le T_f}))$ provides some rare event probability estimate. For example, to obtain the rare event probability estimate that the $T$-time averaged observable exceeds some threshold $a$, we may rewrite eq. \ref{eq.9} as,
\begin{align}\label{eq.10}
&\nonumber E_0\bigg(1_{\{\frac{1}{T} \int_{jT}^{(j+1)T}\phi(x(t)) dt>a|0\le j\le \lfloor T_f/T\rfloor\}}(\phi(x(t)))\bigg)\\ &\sim \frac{1}{N} \sum_{n=1}^N E\bigg(1_{\{\frac{1}{T} \int_{jT}^{(j+1)T}\hat{\phi}(x_n(t)) dt>a|0\le j\le \lfloor T_f/T\rfloor\}}(\hat{\phi}(x_n(t)))\bigg)\cdot e^{-V(\int_0^{T_f} \hat{\phi}(x_n)(t) dt)}\prod_{i=1}^{T_f/\tau} Z_i
\end{align}

A consequence of eq. \ref{eq.10} is that rare event probabilities $P(\Psi\circ\phi(x(t))>a)$ for any functional $\Psi$ of the observed trajectory  $\phi(x(t))$ can be calculated in a similar way.

Hence, rare event probabilities for longer time-averages can be estimated at no further computational expense. Different observables are 
considered in the next section.  We end by remarking that a natural choice is to take $V(x) = Cx$, if the rare event consists of exceedance
of a certain level.

\section{Numerical Results}
 IS algorithms hinge on their ability to shift the sampling distribution of a system to increase the probability of the rare event. They open the possibility of  reducing numerical cost while providing a more (or similarly) accurate estimate over a brute force method. Shifting of the sampling distribution relies on a convergence assumption to hold for a sufficiently large number $N$ of initial particles. In~\cite{Bouchet_Wouters} it is  shown in certain models that the relative error (also a quantity relying on the number of initial particles $N$) is smaller for tail probability estimates obtained from IS methods  if the shift  is chosen optimally for a specific threshold. For a set of thresholds $a_k$, statistics on tail probabilities and return time estimates 
may be obtained by averaging over a set of trials, as in~\cite{Ragone_Wouters_Bouchet}. However, this requirement adds to the true numerical cost of the IS methods.  Optimal values of a shift for any given threshold usually cannot be determined a priori. Moreover, the magnitude of a shift in the sampling distribution cannot be chosen arbitrarily because of its heavy dependence on the choice of observable, system and initial conditions. This dependence limits the algorithm in practice to smaller shift choices, larger errors and hence, lower reliable return-time estimates.

We compare numerical results from two well-known IS methods (GPA and GKLT) with  GEV and MC under true numerical costs of obtaining statistical estimates for sequences of thresholds. In implementation of IS methods, we choose shifting values as large as possible to obtain accurate return-time estimates and illustrate the problems that occur with dependence on initial conditions. Following recent literature, we use the Ornstein-Uhlenbeck process as a benchmark for our work and expand to the more complex Lorenz '96 and PlaSim model. In all systems, we find  that the GEV outperforms GPA, GKLT and MC under the same numerical cost.

\subsection{The Generalized Extreme Value (GEV) Model for Numerical Comparison}
\subsubsection{GEV Model for Comparison to GPA Tail Estimates}
Since the GPA algorithm considers only the distribution of end particles, tail probability estimates of a trajectory $X_t$ are provided at a sampling rate of $T_f$ intervals denoted $P(X_{T_f}>a_k)$ for a sequence of thresholds $a_k$. Recall that in the case of considering an observable under the dynamics, $X_t$ can be seen as the random variable $X_t = \phi(x_t)$ where $x_t$ is the trajectory under the dynamics at time $t$. To compare across methods, we use the same sampling rate for MC brute force and GEV modeling. Following standard literature, we may choose to consider one long trajectory $X_t$ of length $\hat{N}\cdot T_f$, so that we obtain $\hat{N}$ samples taken at $T_f$ intervals of $X_t$. From here, we define the subsequence of $X_t$ taken at the sampling rate $T_f$ to be $X_{\hat{j},T_f}$ for $\hat{j}=1,\cdots,\hat{N}$. We may then define the block maxima over blocks of length $m$ taken over our subsequence $X_{i,T_f}$ by,
\begin{equation*}
M_{\ell,m} = max_{\ell m \le i\le (\ell+1) m}~X_{i,T_f}
\end{equation*}
such that the number of total block maxima is $\lfloor\hat{N}/m\rfloor$ and $\ell = 1,\cdots,\lfloor\hat{N}/m\rfloor$ and $m$ is chosen at a length that ensures convergence of the block maxima. For the purposes of this paper, $m = 10$ and $100$ were checked with $m$ chosen as the value providing the best fit to the control. 

\begin{figure}[h!]
	\centering
	\includegraphics[width=0.5\textwidth]{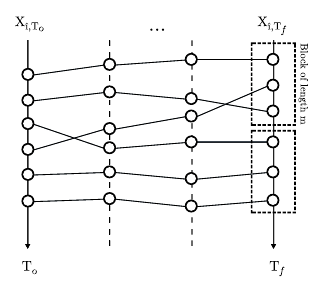}
	\caption{Illustration of the block maxima for GEV to GPA comparison. Many trajectories are run under the dynamics up to the sampling time $T_f$ and the final values are used to form the block maxima (indicated by dashed boxes).\label{fig.GPAGEV}}
\end{figure}

Another option is to run many, say $\hat{N}$ again, trajectories $X_{\hat{i},t}$ for $\hat{i} = 1,\cdots, \hat{N}$ up to time $T_f$. We denote the sequence of end particles $X_{\hat{i},T_f}$ so that $X_{\hat{i},T_f}$ coincides with the appropriate fix sampling rate $T_f$ for each $\hat{i}$. Then, we may define the block maxima over blocks of length $m$ by,
\begin{equation*}
M_{\ell,m} = max_{\ell m \le \hat{i}\le (\ell+1) m}~X_{\hat{i},T_f}
\end{equation*}
so that once again, $\ell = 1,\cdots,\lfloor\hat{N}/m\rfloor$ and the total number of block maxima is $\lfloor\hat{N}/m\rfloor$. In both cases, the distribution of $M_{\ell,m}$ is theoretically the same, however we choose the latter to lower numerical error which builds over long trajectories. An illustration of how the maxima are defined and their relationship to the GPA algorithm outcome can be seen in fig. \ref{fig.GPAGEV}.

Classical results for fitting a GEV to the sequence of block maxima $M_{\ell,m}$ require the sequence $X_{\hat{i},T_f}$ to be independent and stationary. The choice of $T_f>>\tau$ ensures that samples taken at $T_f$ intervals are nearly independent. We may fit the generalized extreme value (GEV) distribution $G(x)$ to the sequence $M_{\ell,m}$ by maximum likelihood estimation of the shape, $\zeta$, scale $\sigma$, and location $\mu$ parameters \cite[Section 3.3.2]{Coles}. Independence assumptions on the sequence $X_{\hat{i},T_f}$ allows for reversibility of the probability estimates of the $m$-block maxima by the following relationship \cite[Section 3.1.1]{Coles},
\begin{equation*}
G(x) = P(M_{\ell,m}\le x) \approx (F(x))^{m}
\end{equation*}
where $G(x)$ is the GEV of the $m$-block maxima estimated by maximum likelihood estimation and $F(x)$ is the c.d.f. of the trajectory $X_t$ sampled at a rate of $T_f$ intervals. Hence,
\begin{equation}\label{ind}
P(X_{T_f}>x) \approx 1-\sqrt[m]{G(x)}
\end{equation}

In the event that independence of $X_{\hat{i},T_f}$ cannot be established, the dependence conditions $D_2$ and $D'$ allow for convergence of the sequence of $m$-block maxima to a GEV distribution.


\subsubsection{GEV Model for Comparison to GKLT Tail Estimates}
In the GKLT algorithm, we consider the distribution of the $T$-time averages created from the $N$-backward reconstructed trajectories $X_{n,t}$. That is, we consider the probability $P(A_{T}>a_k)$  that the $T$-time average, $A_{T} = \frac{1}{T}\int_{0}^{T} X(t)~dt$ is greater than some threshold (or sequence of thresholds) $a_k$. Recall that $X_{n,t} = \phi(x_n(t))$ is some realization of a trajectory under the dynamics equipped with an observable $\phi$. We run $\hat{N}$ trajectories under the dynamics up to time $T_f$ and denote this sequence as $X_{\hat{i},t}$ for $0\le t\le T_f$ and $\hat{i} = 1,\cdots,\hat{N}$. Then the sequence of (non-overlapping) $T$-time averages created from the set of trajectories $X_{\hat{i},t}$ is defined as,
\begin{equation*}
A_{T,\hat{i},j} = \frac{1}{T} \int_{jT}^{(j+1)T} X_{\hat{i},t}~dt
\end{equation*}
for $j = 1,\cdots,\lfloor T_f-T\rfloor$. For each fixed $j$, we define the sequence of maxima taken over blocks of length $m$
\[
M_{h,j,m} = \max_{{h} m\le \hat{i}\le ({h}+1)m}~ A_{T,\hat{i},j}
\]
for $h = 1,\cdots, \lfloor\lfloor T_f-T\rfloor/m\rfloor$ so that we have $\lfloor\lfloor T_f-T\rfloor/m\rfloor\cdot\hat{N}$ number of maxima in total. Defining the maxima over trajectories for every fixed time step $j$, rather than over time steps of a single (long) realization, allows us to keep the integration time small and minimize numerical error. Following previous logic, we may also choose to consider one long trajectory $X_t$, break it up into a sequence of non-overlapping $T$-time averages, and consider the sequence of maxima taken over blocks of length $m$ taken from this long sequence of averages. Once again, we note that $T\ge\tau$ is chosen so that the sequence of averages is roughly independent. Hence, the GEV $G(x)$ can be fit by maximum likelihood estimation to the sequence $M_{h,j,m}$. The independence of the sequence of $T$-time averages allows for reversibility of the probability estimates of the $m$-block maxima by,

\begin{equation*}
G(x) = P(M_{h,j,m}\le x) \approx (F(x))^m
\end{equation*}
where $G(x)$ is the maximum likelihood estimate for the GEV model of the sequence of $m$-block maxima $M_{h,j,m}$ and $F(x)$ is the c.d.f. of the sequence of $T$-time averages taken from the trajectory $X_t$. Hence,
\begin{equation}
P(A_T>x) \approx 1-\sqrt[m]{G(x)}
\end{equation}

\begin{figure}[h!]
	\centering
	\includegraphics[width=0.5\textwidth]{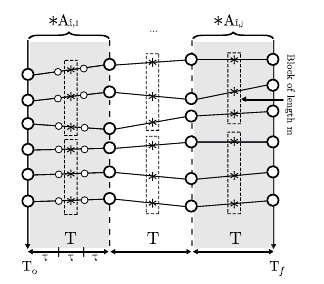}
	\caption{Illustration of the block maxima for GEV to GKLT comparison. Many trajectories are run under the dynamics up to time $T_f$. $T$-time average sequences are calculated from the trajectories. For each fixed time step $j$, the block maxima (indicated by dashed boxes) are calculated. The $\tau$ interval is shown here to emphasize its difference to $T$ and does \textbf{not} represent any weighting done to trajectories used in the GEV model. \label{fig.GKLTGEV}}
\end{figure}

An illustration of how the block maxima in estimating the GEV are defined in terms of the sequence of $T$-time average trajectories for comparison to the GKLT algorithm can be found in fig. \ref{fig.GKLTGEV}.

\subsubsection{Return Time Curves}

We consider a long trajectory $X_t$ such that $X_t$ is sampled for over threshold probability estimates at time $T_f\ge \tau$ and a rare event threshold $a$ such that $X_t<a$ for most times $t$. We define the return time $r(a)$ as the average waiting time between two statistically independent events exceeding the value $a$.

Following the logic in \cite{Ragone_Wouters_Bouchet} we divide the sequence $X_t$ into pieces of duration $\Delta T$ and define $a_k = \max\{X_t|(k-1)\Delta T\le t\le k\Delta T\}$ and $s_k(a) = 1$ if $a_k>a$ and 0 otherwise. Then the number of exceedances of the maxima $a_k$ over threshold $a$ can be approximated by a Poisson process with rate $\lambda(a) = 1/r(a)$. Using the return time c.d.f. $F_T^{-1}$ for the Poisson process, we have

\[
F_T^{-1}(\frac{1}{K}\sum_{k=1}^K s_k(a)) = \frac{-\log(1-\frac{1}{K}\sum_{k=1}^K s_k(a))}{\lambda(a)}
\]

where $\frac{1}{K}\sum_{k=1}^{K} s_k(a) = F_T(\Delta T)$ is the probability of observing a return of the maxima $a_k$ above threshold $a$ in $\Delta T$ time steps. For any $a_k$ we have an associated probability $p_k$. We denote the reordering of this sequence $(\hat{a}_k,\hat{p}_k)$ such that $\hat{a}_1\ge\hat{a}_2\ge\cdots\ge\hat{a}_K$. Then the return time is given by,

\begin{equation}\label{rtn}
r(\hat{a}_k) = -\frac{1}{\log(1-\sum_{k=m}^K \hat{p}_m)}
\end{equation}
where $\sum_{m=k}^K \hat{p}_m$ gives an approximation of the probability of exceeding threshold $\hat{a}_k$.

Return time plots estimated from outcomes of importance sampling methods are the result of first averaging return time estimates over a number of experiments for each $C$, then averaging over all $C$-return time plots. See fig. (\ref{fig.RT}) for an illustration. Only those return times corresponding to threshold values that fall within $1/2$ standard deviation of the tilted distribution are used in this averaging. For the remainder of this paper, the term \textit{experiment} will be used to describe a single run of an importance sampling algorithm.


\subsection{Ornstein-Uhlenbeck Process} \label{OU}
The Ornstein-Uhlenbeck process given by,
\[
dx = -\lambda x dt +\sigma d\mathcal{W}
\]
is a nice toy-example for importance sampling application because it is simple to implement, has low numerical cost, the distribution of position $x$ is approximately  Gaussian, and it's correlations decay exponentially. We use this process with $\lambda=1$ and $\sigma=1$ as a benchmark for the following numerical investigation.

\subsubsection{GKLT}

The GKLT importance sampling algorithm is performed on the Ornstein-Uhlenbeck process with $N=100$ initial trajectories, resampling time $\tau=0.1$, and a total integration time of $T_f=2.0$. Here, the observable of interest is the position. At each time step of the algorithm, a new value of noise $\mathcal{W}$ is sampled from the standard normal distribution for each cloned trajectory to ensure divergence of the clones. Time average trajectories are calculated by averaging the $N=100$ backward-reconstructed trajectories over time-windows of length $T=0.25$ with step size equal to $T$ so that no window has overlapping values. 

Above threshold probabilities of the $T$-time average position $P(A_T>a_k)$ where $A_T = \frac{1}{T}\int_{0}^{T} x(t)~dt$ are estimated for $C = [0.01, 0.03, 0.05, 0.07]$. We define the sequence of $T$-time averages obtained from realizations $\hat{\phi}(x_n(t))$ of the $N$-backward reconstructed trajectories as,
\begin{equation}\label{obs1}
A_{n,j}  = \frac{1}{T} \int_{jT}^{(j+1)T} \hat{\phi}(x_n(t)) dt,
\end{equation}
for $j = 1,\cdots,\lfloor T_f/T\rfloor$. Then the probability estimate for $P(A_T(t)>a)$ above a threshold $a$ from eq. \ref{eq.10} is given as,
\[
E_0(1(x(t))_{\{A_T>a|0\le t\le T_f-T\}}) \sim \frac{1}{N} \sum_{n=1}^N E(1(\hat{\phi}(x_n(t))_{\{A_{n,j}>a|j = 1\cdots \cdot\lfloor T_f/T\rfloor \}})e^{-C(\int_0^{T_f} \hat{\phi}(x_n(t)) dt)}\prod_{i=1}^{T_f/\tau} Z_i.
\]
This approach results in a unique probability estimate for each predefined threshold $a$.

\begin{figure}
	\includegraphics[width=\textwidth]{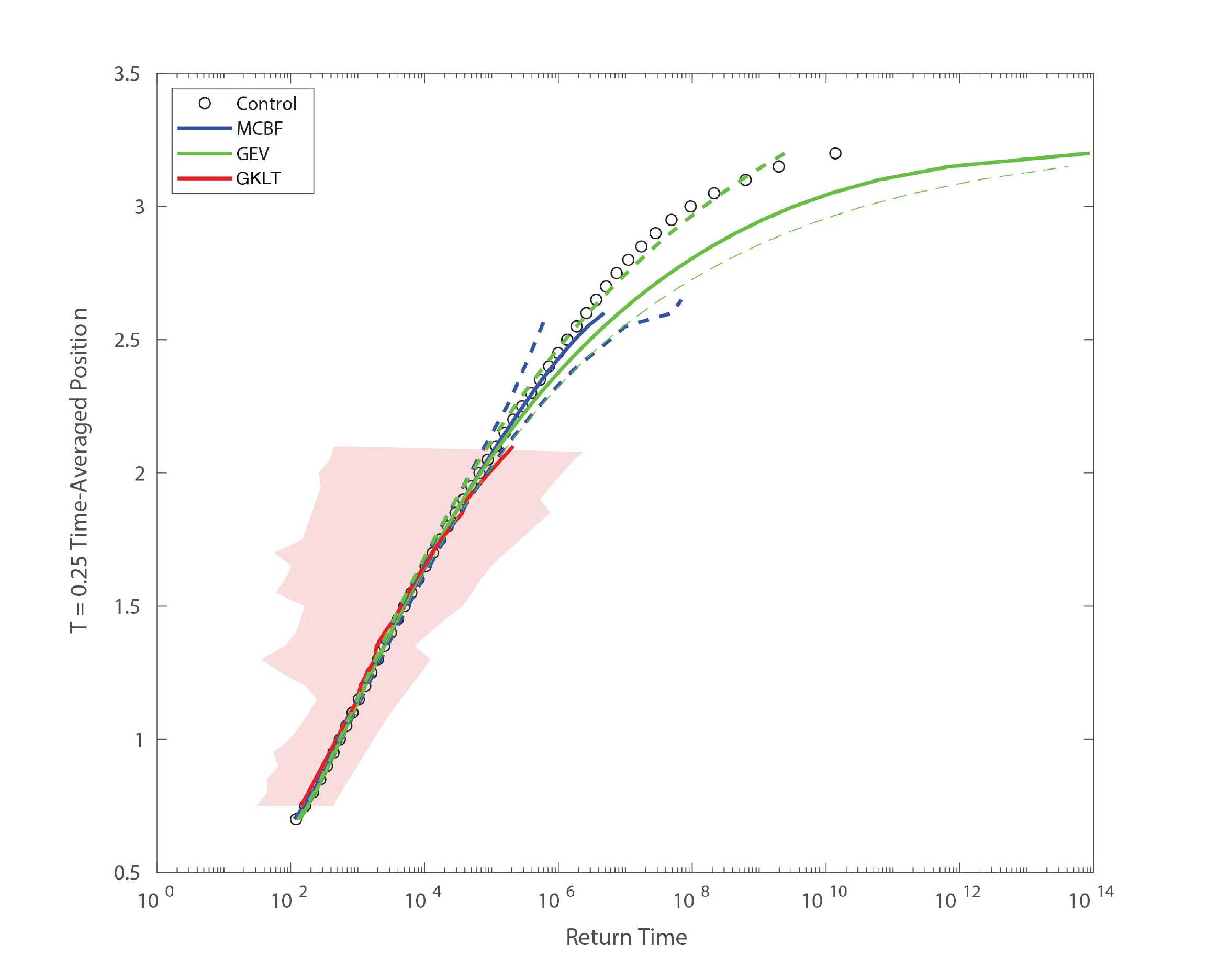}
	\caption{Return time estimates for the Ornstein Uhlenbeck process time average observable using GKLT for 4 different $C$ values and 100 experiments, GEV, and Monte Carlo brute forces methods with numerical cost $4\cdot 10^4$. Relative error curves for MC brute force and GEV estimates are represented by dashed lines. Relative error estimated by 100 experiments of the GKLT process is represented by the shaded red region.\label{fig.GKLT_OU_B}}
\end{figure}

Return times are estimated for each value $C$ and sequence of thresholds $a_k$ by eq. \ref{rtn} resulting in four return time curves. We perform 100 experiments under these conditions for a total of 400 return time curves and average to obtain the result shown in fig. (\ref{fig.GKLT_OU_B}). This process is illustrated in fig (\ref{fig.RT}). The total numerical cost for this estimate is $4\cdot 10^4$. Monte Carlo (MC) brute force and generalized extreme value (GEV) (eq. \ref{ind}) probability estimates are obtained through numerical costs of the same order. We find that GEV and MC brute force methods outperform GKLT by providing estimates of return times longer than $1\cdot 10^6$.

 Another option is to define the sequence corresponding to the maximum $T$-time average quantity of a single realization $\hat{\phi}(x_n)$ given by,

\begin{equation}\label{obs2}
 a_n(T) = \max_{1\le j\le \lfloor T_f/T\rfloor} \frac{1}{T}\int_{jT}^{(j+1)T} \hat{\phi}(x_n(t)) dt.
\end{equation}

\begin{figure}
	\includegraphics[width=\textwidth]{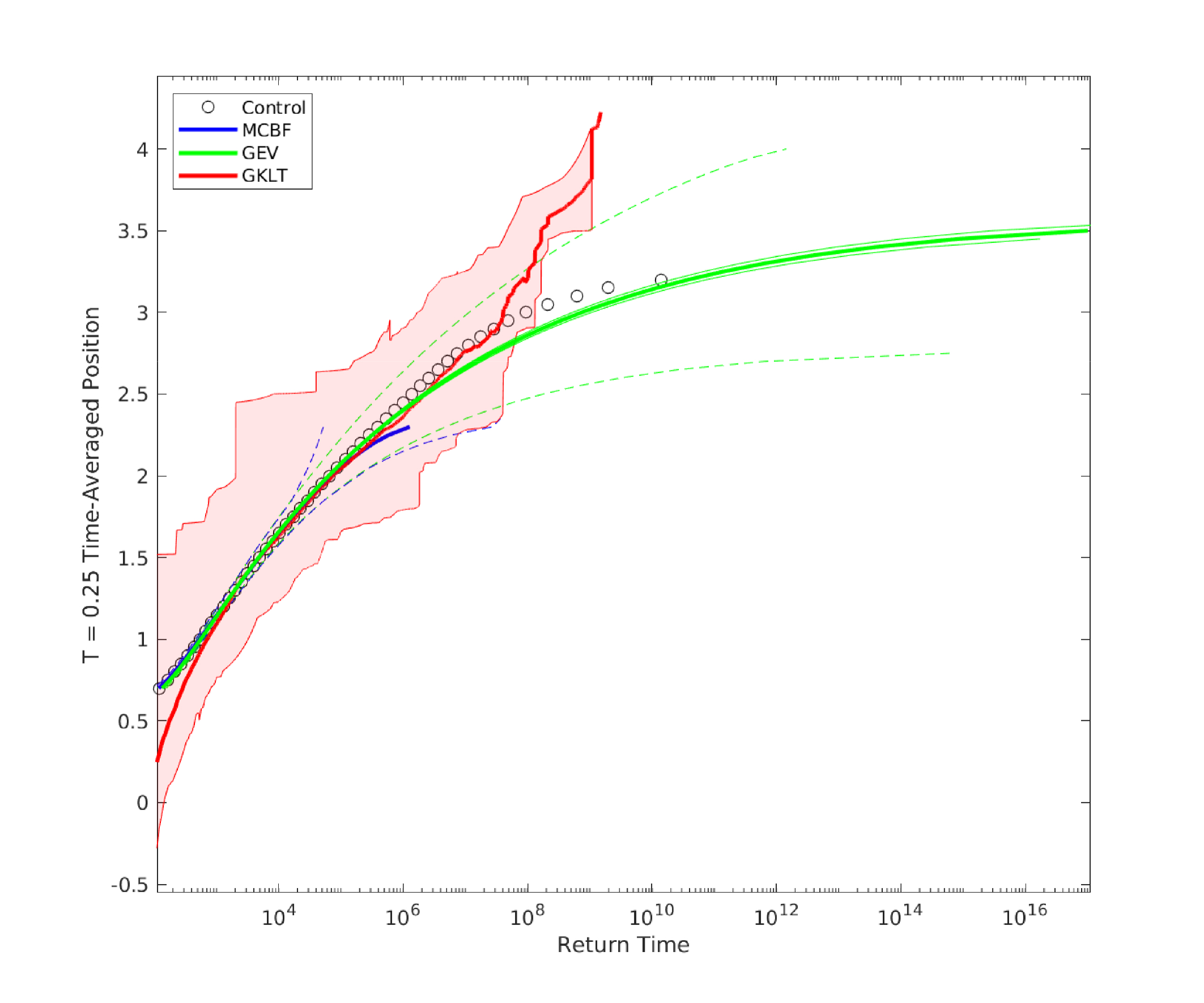}
	\caption{Return time estimates \textbf{from the sequence of maxima taken over each trajectory} for the Ornstein Uhlenbeck process time average observable using GKLT for 4 different $C$ values and 100 experiments, GEV, and Monte Carlo brute forces methods with numerical cost $4\cdot 10^4$. Relative error estimates for GEV and MC methods (dashed lines) and GKLT (red region) are estimated from 100 experiments.\label{fig.GKLT_OU}}
\end{figure}

This results in a sequence of maximum thresholds $a_n(T)$, one per each realization of $\hat{\phi}(x_n(t))$. For each threshold $a_n(T)$, there exists an associated probability estimate,
\[
p_n = \frac{1}{N} e^{-C\int_{0}^{T_f}\hat{\phi}(x_n(t)) dt}\prod_{i=1}^{T_f/\tau}Z_i,
\]
which is the result of plugging the threshold values of eq. \ref{obs2} into eq. \ref{eq.10} and noting that,
\[
E(1(x_n(t))_{\{\frac{1}{T}\int_{jT}^{(j+1)T}\hat{\phi}(x_n(t))dt>a_n(t,T)|0\le t\le T_f-T\}}) = 1.
\]
The sequence $(a_n(T),p_n)$ for $1\le n\le N$ is then reordered for decreasing values of $a_n$. We denote the ranked sequence $(\hat{a}_n(t),\hat{p}_n)$ where $\hat{a}_1\ge \hat{a}_2\ge\cdots\ge \hat{a}_N$ and associate a return time $r(\hat{a}_n)$ defined by eq. \ref{rtn} using the reordered sequence $\hat{p}_n$. We refer to \cite{Ragone_Wouters_Bouchet} for more details on this approach. Return time curves are then obtained by linearly interpolating the pair $(\hat{a}_n(T),r_n(\hat{a}_n))$ over an equal spaced vector of return times. GKLT is run with the same initial conditions as stated above. We refer to fig. (\ref{fig.GKLT_OU}) for this discussion. Choosing to calculate return time curves in this way allows for estimates of longer times; however, this tends to be at the expense of accuracy. Equation. \ref{obs1} allows for more control over the choice of range of thresholds included from the shifted distribution.

GEV and MC estimates are obtained through numerical costs of the same order. Deviation statistics for GKLT, GEV, and MC methods, represented by dashed lines in fig. (\ref{fig.GKLT_OU}), are calculated by finding the minimum and maximum deviation in 100 experiments. Solid lines about the GEV represent the 95\% confidence intervals coming from the likelihood function for the GEV estimated from the corresponding MC simulation. We compare all results against a long control run of order $1\cdot 10^6$. We find that GEV and GKLT methods provide more accurate estimates of return times longer than $1\cdot 10^5$ compared to the MC method. Moreover, the GEV outperforms the GKLT algorithm by providing surprisingly accurate return time estimates with smaller deviation for all thresholds except in a small fraction of cases.

\begin{figure}
	\includegraphics[width=\textwidth]{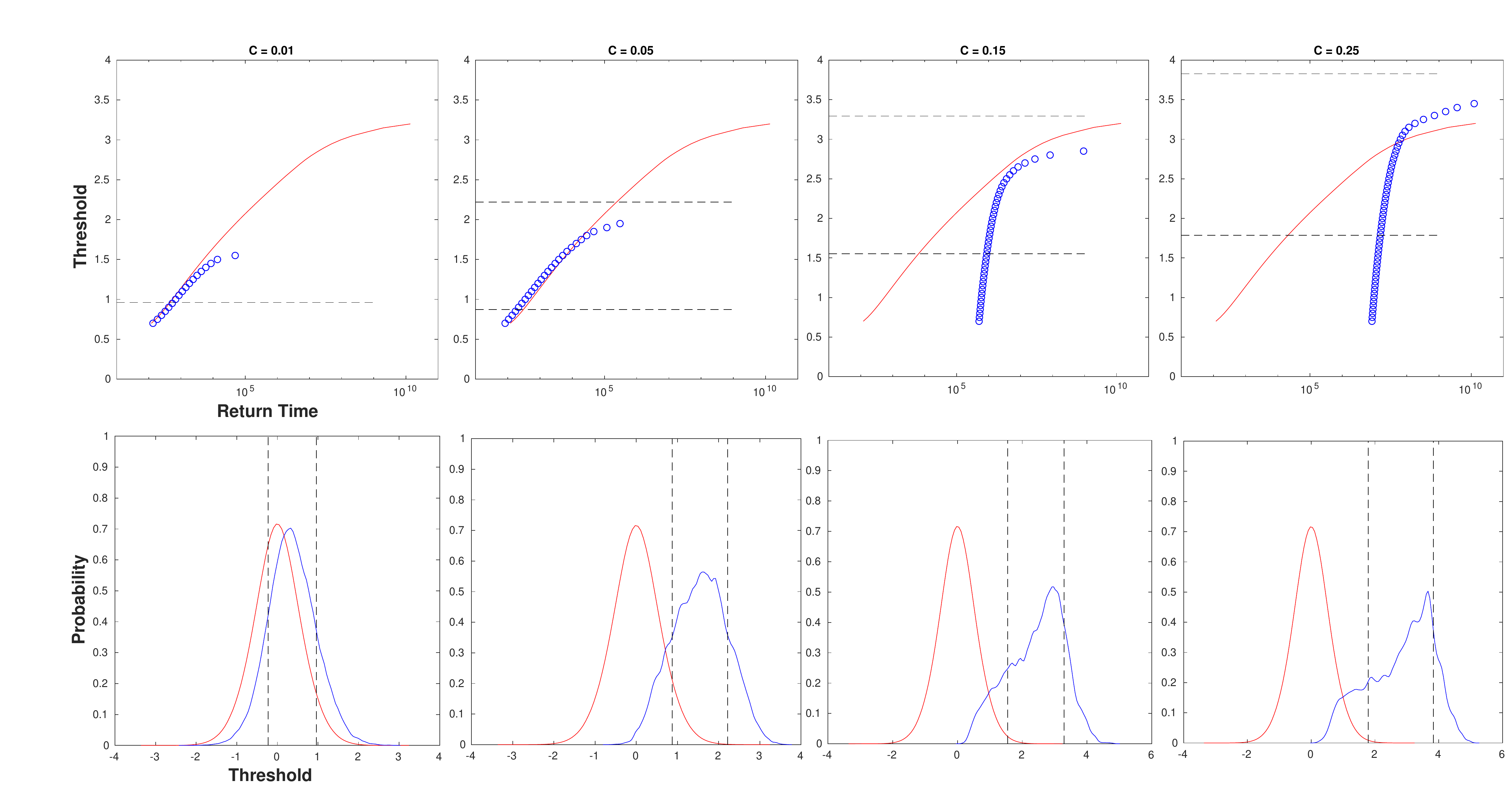}
	\caption{Return time estimates for the Ornstein Uhlenbeck process time average observable illustrating the breakdown of the distributions for large values of $C$.\label{fig.RTB}}
\end{figure}

A possible  explanation for the poor performance of the GKLT algorithm comes from the fact that the tilting coefficient $C$ cannot be chosen arbitrarily large to obtain longer return time estimates without some change in the initial conditions (e.g. integration time, number of starting trajectories). Large choices of $C$ result in a lower number of parent trajectories (as many copies are killed) which causes the tilted distribution to breakdown fig. (\ref{fig.RTB}). This breakdown results in increasingly inaccurate return time estimates, even for thresholds sitting close to  the center of the tilted distribution.

\subsubsection{GPA}

\begin{figure}
	\includegraphics[width=\textwidth]{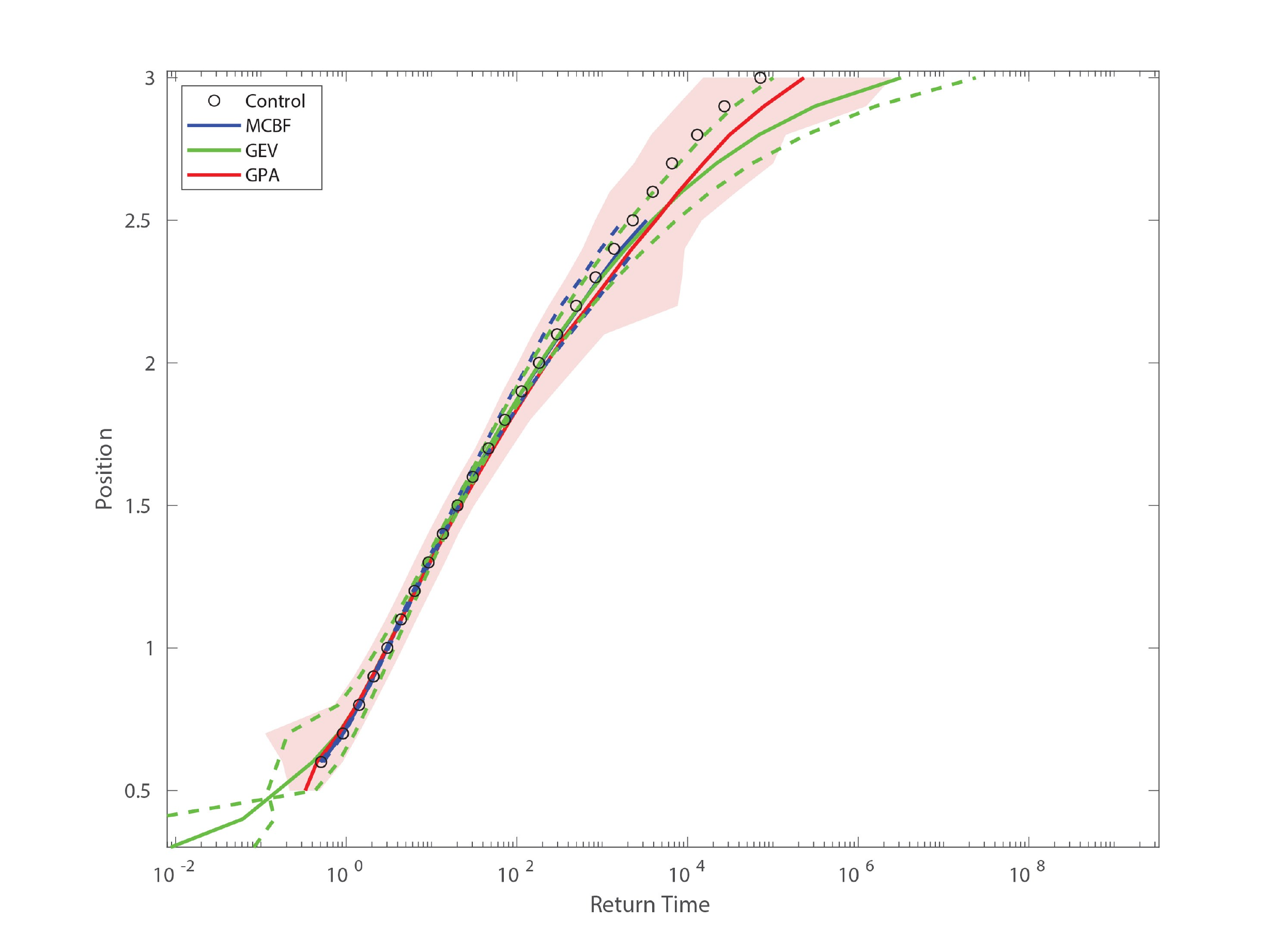}
	\caption{Return time estimates for the Ornstein Uhlenbeck process using GPA for 3 different $C$ values estimated over 10 experiments, GEV, and Monte Carlo brute forces methods with numerical cost $3\cdot 10^3$. Relative error estimates for GEV amd MC methods (dashed lines) and GPA (red region) are estimated from 10 experiments.\label{fig.GPA_OU}}
\end{figure}

The GPA importance sampling algorithm is performed with $N=100$ starting trajectories, resampling time $\tau=0.1$, and a total integration time of $T_f=2.0$. The final trajectories $X_{n,T_f}$ from GPA with tilting constants $C = [2,3,4]$ are used to estimate the above threshold probabilities $P(X_{T_f}>a_k)$ and return time curves. To begin, we perform 10 experiments, with the initial conditions described above, resulting in a total of 30 return time curves (10 experiments for each value of $C$) and average to obtain the result shown in fig. (\ref{fig.GPA_OU}). The total numerical cost for this estimate is $3\cdot 10^3$ compared to the long control run of $1\cdot 10^6$. We find that GPA and GEV methods provide nearly equivalent results fro return times up to $1\cdot 10^4$ with GPA and GEV methods outperform Monte Carlo brute force estimates for return times longer than $1\cdot 10^4$. On average, GPA provides a slightly closer approximation to the control curve than that of the GEV method for longer return times; however, the deviation of this estimate is much larger than that of GEV.

\begin{figure}
	\includegraphics[width=\textwidth]{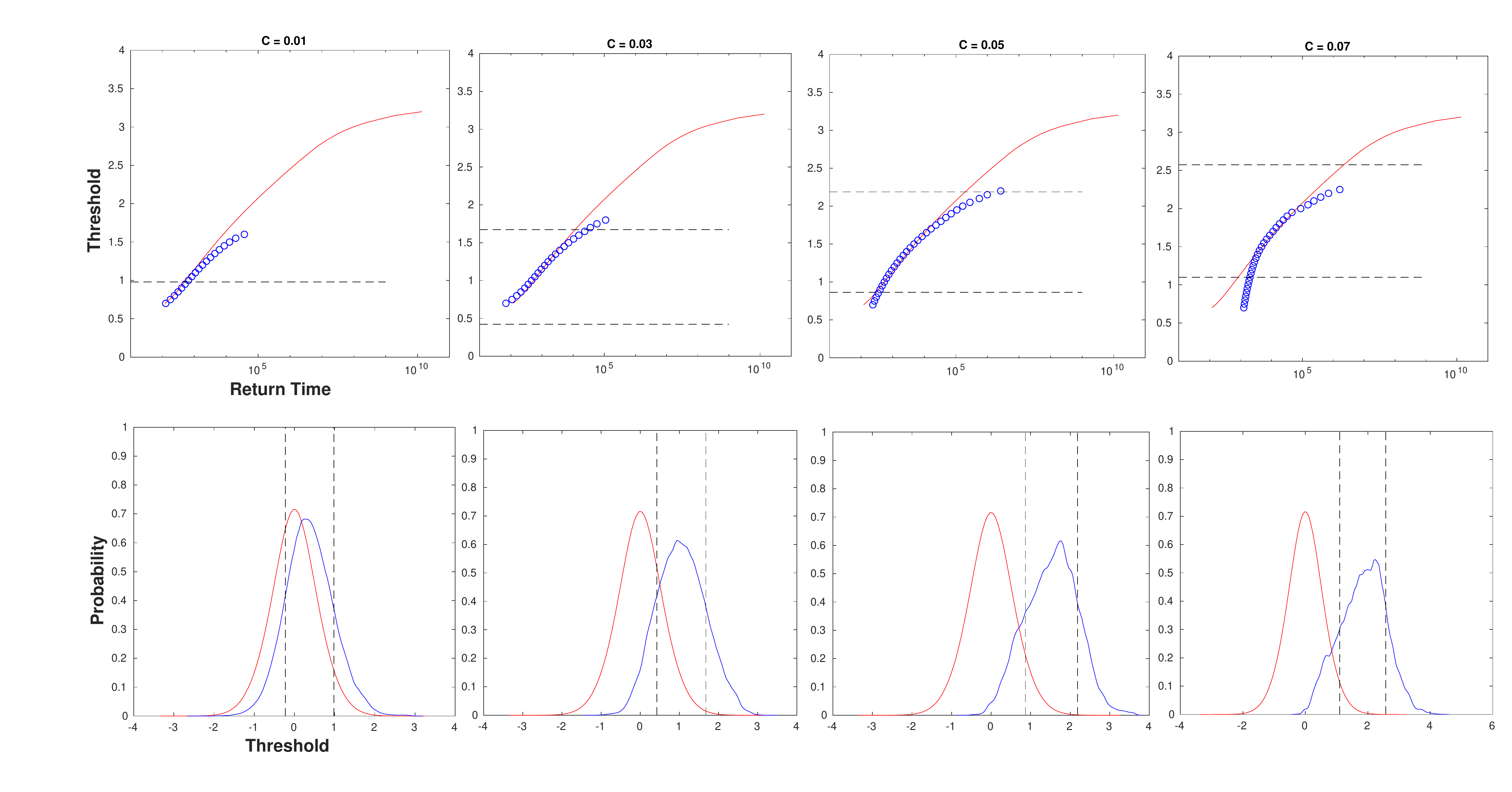}
	\caption{Return time estimates for the Ornstein Uhlenbeck process time average observable illustrating the choice of return time curves after GKLT implementation. \label{fig.RT}}
\end{figure}

\begin{figure}
	\includegraphics[width=\textwidth]{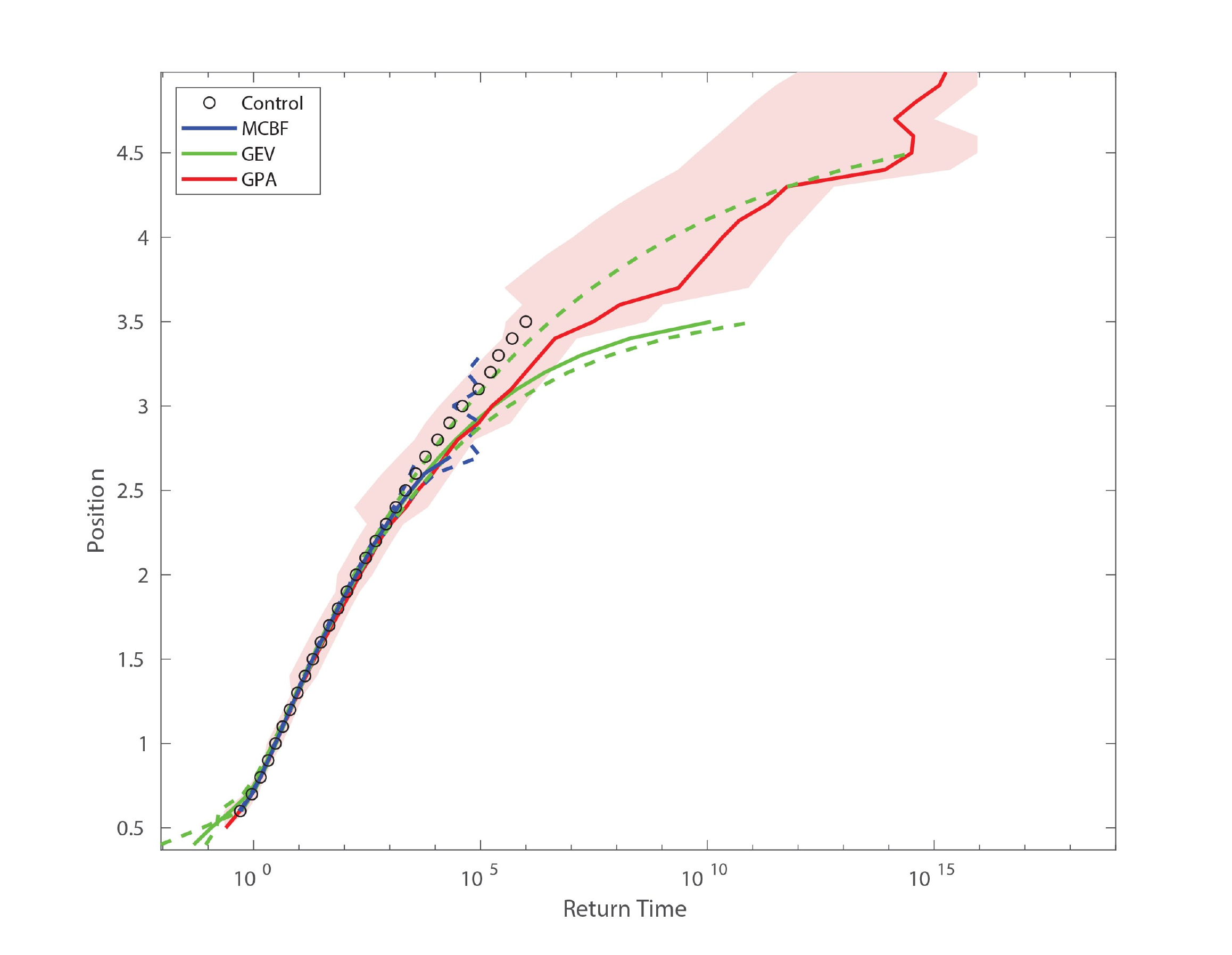}
	\caption{Return time estimates for the Ornstein Uhlenbeck process using GPA for 10 different $C$ values estimated over 30 experiments, GEV, and Monte Carlo brute forces methods with numerical cost $3\cdot 10^3$. Relative error estimates for GEV amd MC methods (dashed lines) and GPA (red region) are estimated from 30 experiments.\label{fig.GPA_OU_C}}
\end{figure}

Next, we consider larger values of $C$ to test whether reliable estimates can be obtained for thresholds exceeding the control run. We run 30 experiments for 10 different values of $C = [1,2,\dots,10]$ under the same initial conditions as stated above for a total numerical cost of $3\cdot 10^4$. We average the resulting return time curves shown in fig. (\ref{fig.RT}) to obtain the final return time plot fig. (\ref{fig.GPA_OU_C}). As seen in the estimates for GKLT, higher values of $C$ with unchanged initial conditions provide less accurate return-time results even for those thresholds which sit at the center (e.g. have the highest probability of occurrence) of the tilted distribution. On the other hand, GEV methods with the same numerical cost of $3\cdot 10^4$ show surprisingly reasonable estimates for return times longer than the control method can provide at numerical costs of $1\cdot 10^6$.

\subsubsection{Relative error estimates.}

\begin{figure}
	\includegraphics[width=\textwidth]{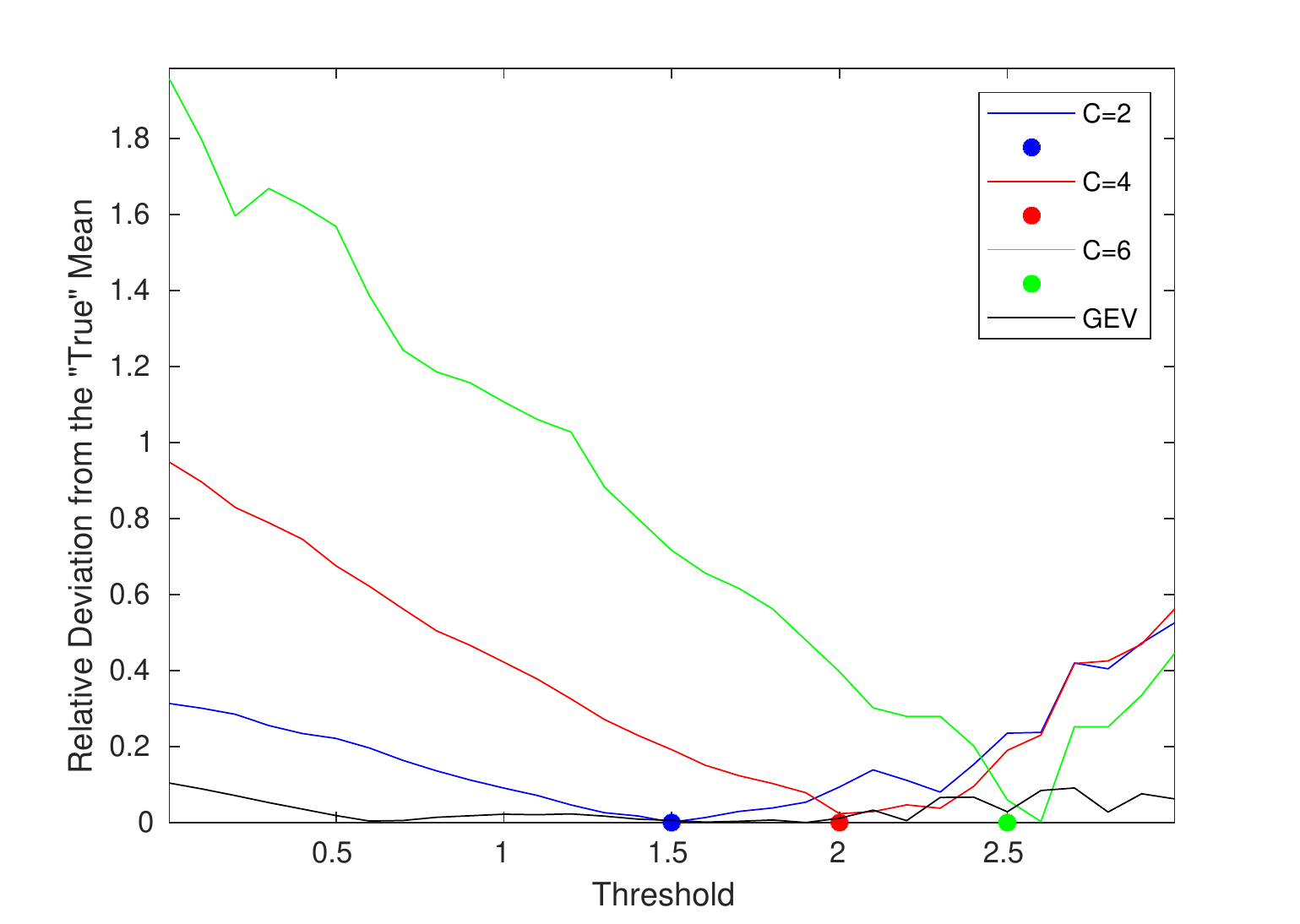}
	\caption{Relative deviation of the estimated mean $\mu(\hat{\gamma_A})$ from $K=100$ runs of GPA with $N=1000$ from the assumed, asymptotic mean $\gamma$. This deviation is only near zero for thresholds whose optimal tilting value $C$ is chosen in the weight function (marked with a $\circ$). Relative deviation of the estimated mean from the GEV method is consistently near zero, suggesting that even though the deviation is larger, the estimate is more reliable. \label{fig.MEAN_RE}}
\end{figure}

We now discuss relative error estimates on return probabilities across GPA, GEV, and MC methods. The relative error is estimated as $\sqrt{\sum_{j=1}^K \frac{1}{K}(\hat{\gamma}-\gamma)^2}/\gamma$ where $\hat{\gamma}$ is the estimate for each of $K=100$ experiments and $\gamma$ is the long control-run estimate. The relative error is essentially the average deviation of the tail probability estimate $\hat{\gamma}$ from the true value $\gamma$ where it is assumed that $\hat{\gamma}$ follows a Gaussian distribution with mean $\gamma$ \cite{Bouchet_Wouters,Garnier_DelMoral} \textit{for a sufficiently large number $N$ of starting particles}. For lower values of $N$, the relative error calculated in this way has an underlying measurement error in the bias that is observed for $\hat{\gamma}$ in lower $N$ values. Although this bias is often considered negligible, the sensitivity of long return times to small deviations in the tail probability estimate suggest otherwise. We first illustrate that the relative error cannot be used reliably for thresholds whose optimal tilting value is not approximately $C$. We calculate an estimate of the mean $\mu(\hat{\gamma}) = \frac{1}{K}\sum_{k=1}^K \hat{\gamma}_k$ for $K=100$ experiments with $N=1000$ and three different values of $C$. Then, we calculate the relative deviation of $\mu(\hat{\gamma})$ from the "true" mean $\gamma$ by $\sqrt{(\mu(\hat{\gamma})-\gamma)^2}/\gamma$ for each value of the threshold. Results in fig. (\ref{fig.MEAN_RE}) show the this deviation is small only for thresholds whose tilting  value $C$ lies near the optimal  value. 

\begin{figure}
	\includegraphics[width=\textwidth]{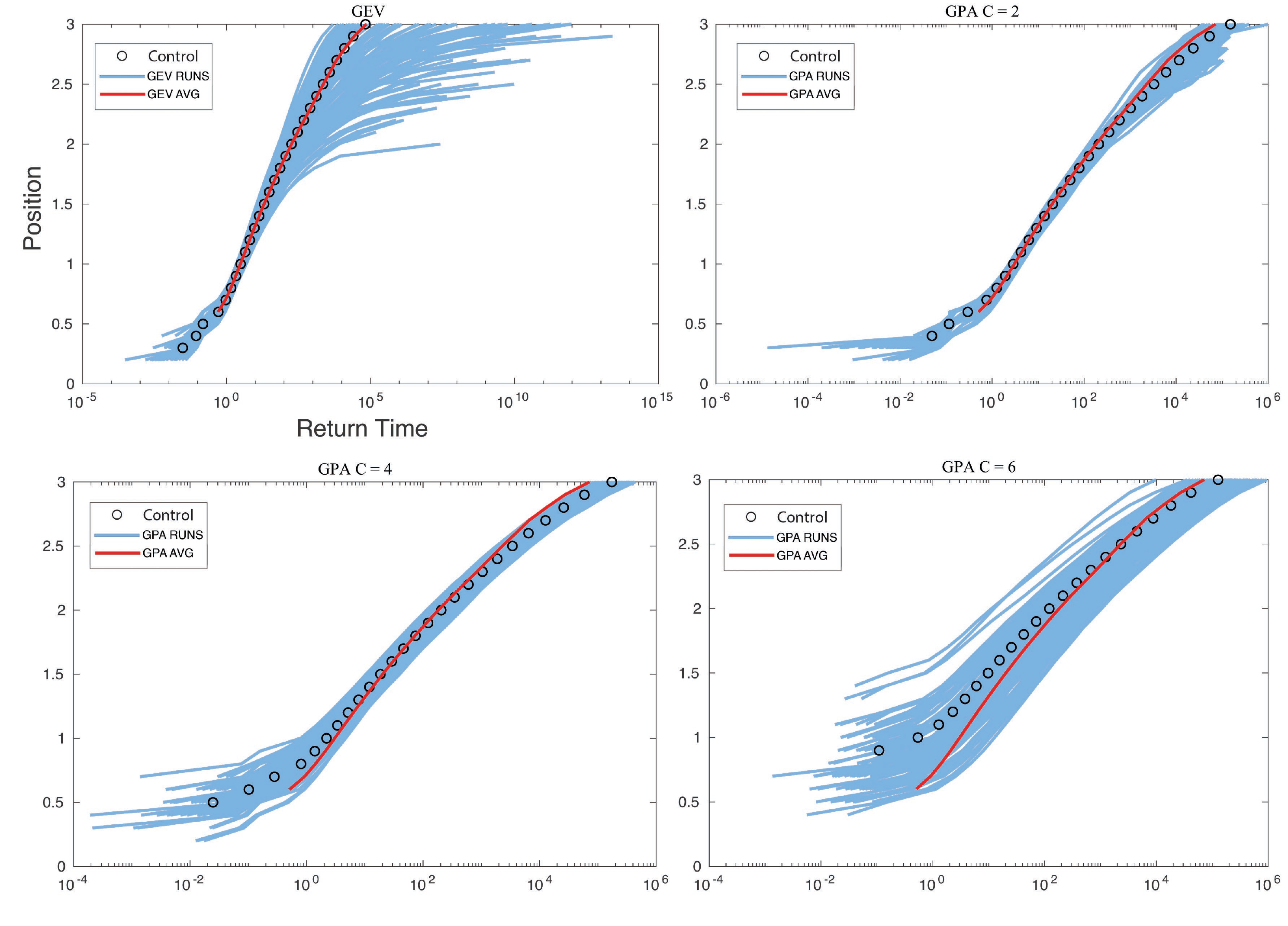}
	\caption{Illustration of the deviation of the return time curves from the control for GEV and 3 different tilting $C$ values of GPA. Notice that the average return time curve (red) for the GEV fits the control (black $\circ$) for all long return times while accurate estimates for GPA only occur near the optimal threshold value.\label{fig.RE_SPREAD}}
\end{figure}

The effects of this deviation can be seen in return time estimates. We calculate the return time curves from 100 experiments of GPA and GEV methods with $N=1000$ fig. (\ref{fig.RE_SPREAD}) Clearly, GEV methods produce a larger standard deviation for return times. Under the assumptions above, the relative error for GEV methods would be larger than that of GPA; however, the mean of the tail probabilities obtained from GEV are nearly exactly   those of the long control run. On the other hand, GPA produces a much smaller standard deviation (relative error) while the mean of the tail probabilities have accurate estimates only near thresholds for which the $C$ value is chosen optimally.

\begin{figure}
	\includegraphics[width=\textwidth]{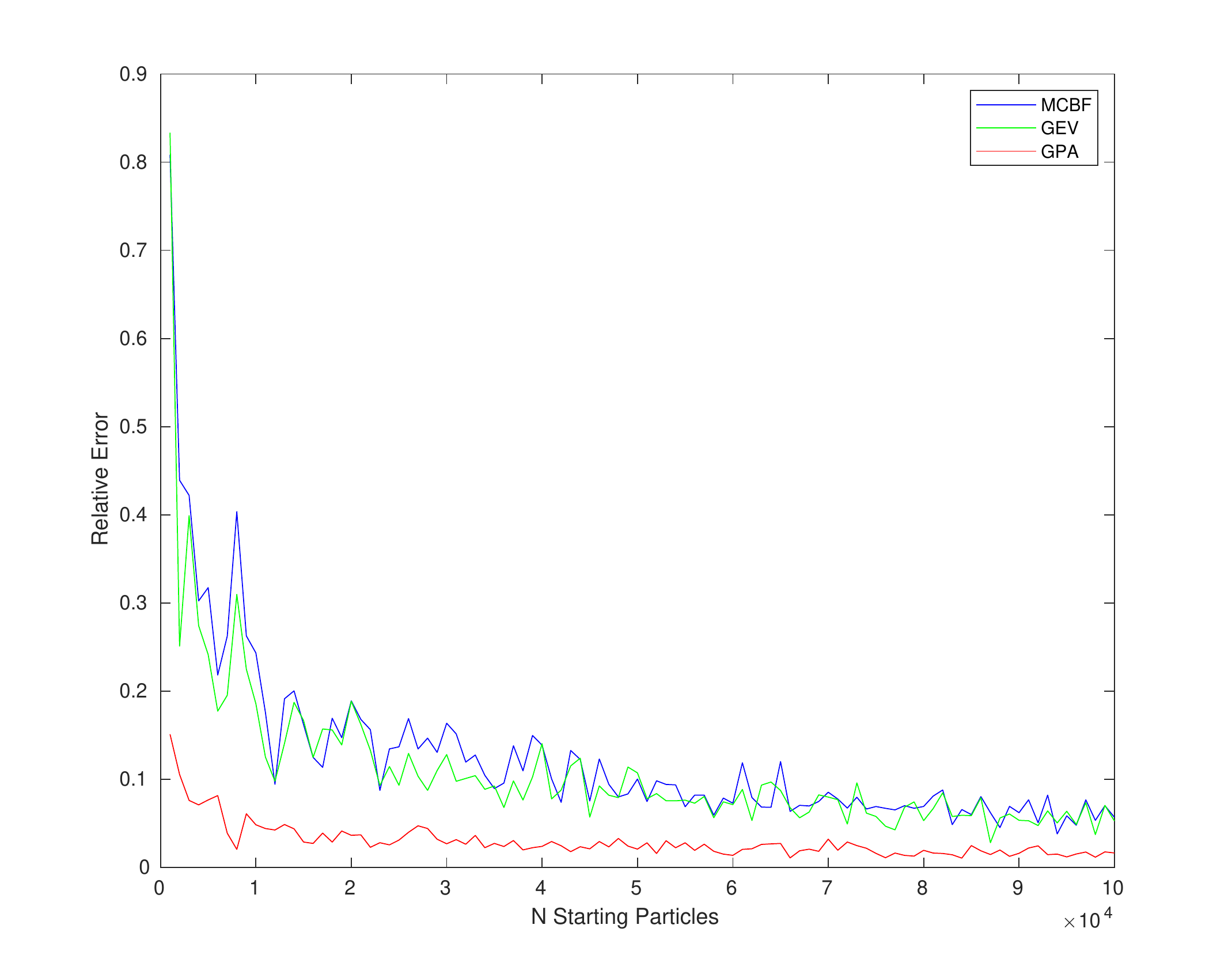}
	\caption{Relative error for MC, GEV, and GPA probability estimates of fixed threshold 2 and corresponding optimal tilting value $C=4$.\label{fig.RE}}
\end{figure}

We remark that for a single threshold and a close to  optimal value of $C$, relative error estimates are reliable and GPA outperforms GEV and MC methods under relative error fig. (\ref{fig.RE}) while providing accurate return time estimates fig. (\ref{fig.RE_SPREAD}). These results are consistent with those of \cite{Bouchet_Wouters}. Interestingly, though not surprisingly, are the results on equivalent relative error for the GEV and MC methods for shorter return times. This equivalence suggests that the advantage of GEV over MC methods comes from its ability to estimate longer return times where MC methods fail to provide results.

\subsection{Lorenz Model}

 The Lorenz 1996 model  consists of $J$ coupled sites $x_l$ on a ring, 
 \[
 \dot{x_l}=x_{l-1}(x_{l+1}-x_{l-2})+ R-x_l
 \]
 $l=0,\ldots,J-1$ where the indices are in $\mathbb{Z}^J$. The parameter $R$ is a forcing term and the dynamics is chaotic for $R\ge 8$~\cite{Lorenz_1996,Lorenz_2005}. The energy 
 $E(x)=\frac{1}{2J}\sum_{l=1}^{J}x_l^2$ is conserved and there is a repelling fixed hyperplane $x_l=R$, $l=0,\ldots,J-1$. The extremes of interest investigated numerically in \cite{Bouchet_Wouters} and in  our preliminary work were tail probabilities of the form $P(E(x(t))>E_t)$. The energy observable on this system has an approximately Gaussian distribution.
 
 \subsubsection{GPA, GEV and MC.}

 The weight function is taken to be the change $\Delta E$ of energy i.e. $E(x(t+1))-E(t)$ for a single time step and from this an exponential weight function  $W=\exp(C\Delta E)$ is constructed, depending on a single parameter $C$ (large $C$ makes tail probabilities greater). For this analysis, we choose $J = 32$ sites and a forcing coefficient $R = 64$.
 
\begin{figure}
	\includegraphics[width=\textwidth]{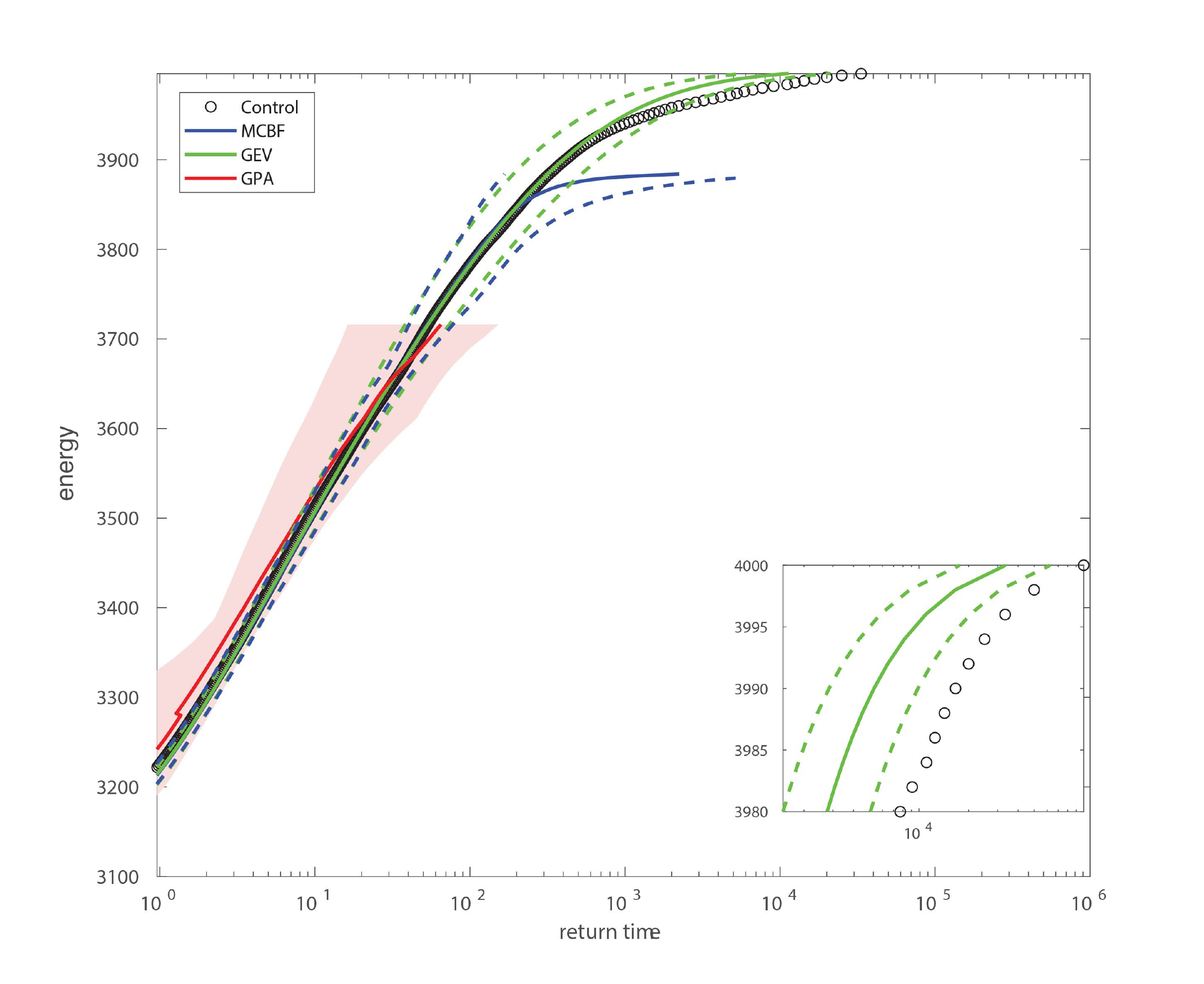}
	\caption{Return time estimates for the Lorenz '96 process using GPA for $C=[3.1\cdot 10^{-3}, 6.4\cdot 10^{-3}]$ estimated over 10 experiments for $N=2000$ starting particles, GEV, and Monte Carlo brute forces methods with numerical cost $4\cdot 10^4$. Relative error estimates for GEV amd MC methods (dashed lines) and GPA (red region) are estimated from 10 experiments. \label{fig.GPA_LORENZ_2}}
\end{figure}

\begin{figure}
	\includegraphics[width=\textwidth]{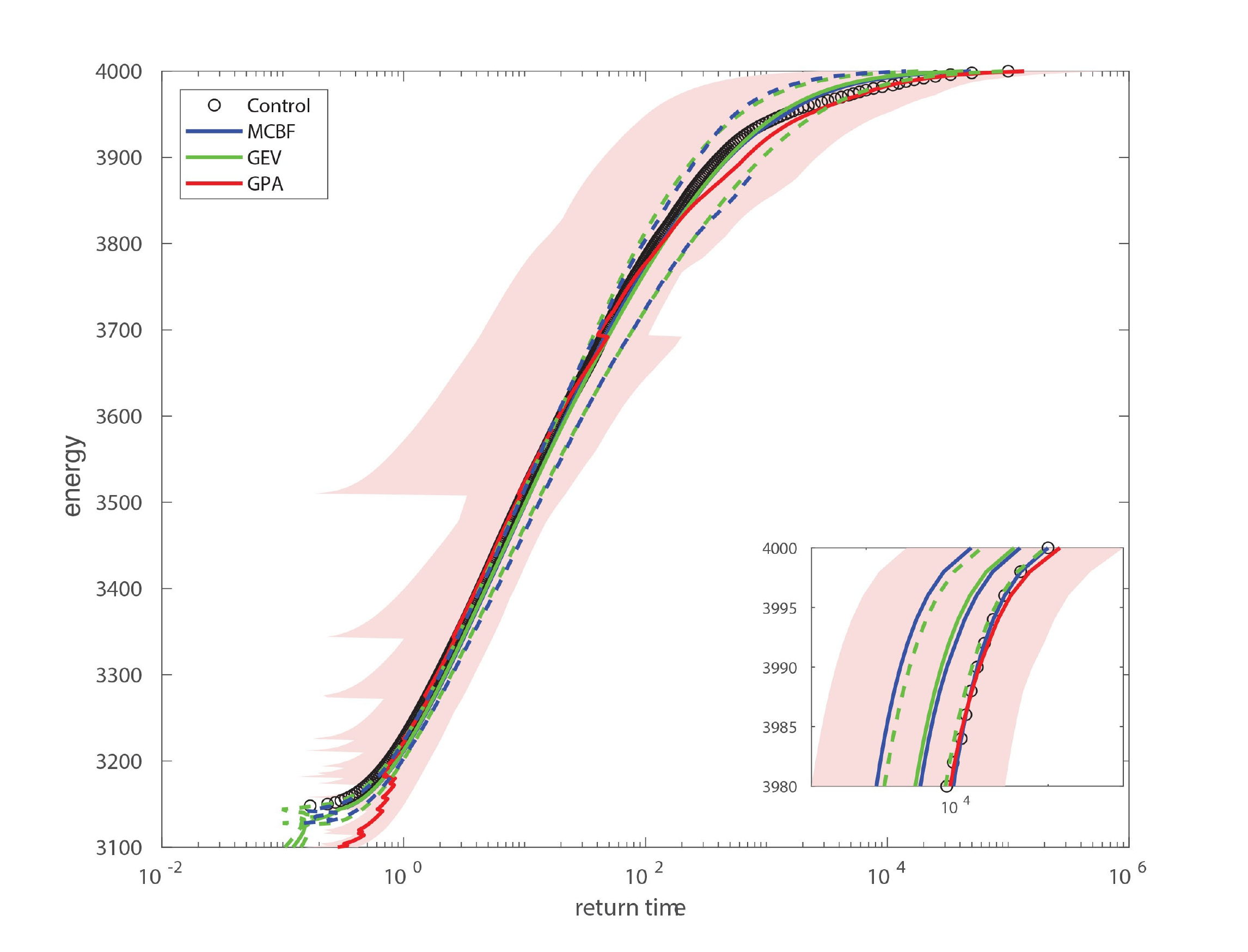}
	\caption{Return time estimates for the Lorenz '96 process using GPA for $C=[3.1\cdot 10^{-3},6.4\cdot 10^{-3}]$ estimated over 10 experiments for $N=5000$ starting particles, GEV, and Monte Carlo brute forces methods with numerical cost $1\cdot 10^5$. Relative error estimates for GEV amd MC methods (dashed lines) and GPA (red region) are estimated from 10 experiments. \label{fig.GPA_LORENZ_1}}
\end{figure}

 The GPA importance sampling algorithm is performed with $N = 2000$ and $5000$ starting trajectories, a resampling time $\tau=0.08$, and a total integration time of $T_f = 1.28$. At each time-step of the algorithm, a random perturbation sampled from $[-\epsilon,\epsilon]$ where $\epsilon = O(10^{-3})$ is added to the clones of the previous iteration to ensure divergence. The final trajectories from GPA with tilting constants $C = [3.2\cdot 10^{-3}, 6.4\cdot 10^{-3}]$ are used to calculate the above threshold probabilities and return time curves. The return time curve is calculated by averaging over 10 experiments. Return time curves from the GEV and MC methods are created from runs of equal numerical cost $4\cdot 10^4,$ and $1\cdot 10^5$, respectively. All estimates are compared to a long control run of $1\cdot 10^6$. For $N = 2000$ initial starting particles both GEV and MC methods outperform GPA by providing more accurate return time estimates for times longer than $1\cdot 10^3$ (fig. \ref{fig.GPA_LORENZ_2}). GPA seems to provide more accurate estimates for returns longer than $1\cdot 10^5$ for $N=5000$; however, the deviation of the averaged return time curve is much larger than that of GEV or MC methods for all thresholds (fig. \ref{fig.GPA_LORENZ_1}). 
 
 The complexity of the Lorenz '96 highlights some of the major pitfalls in GPA. Intuitively, the choice of tilting value $C$ is (roughly) the shift required for center of the distribution of the observable to lie directly over the threshold of interest. The Lorenz system provides an example of the difficulties involved in choosing this tilting value in practice. Similar to the OU system, the underlying dynamics of the Lorenz system equipped with the energy observable cause a breakdown in the shifted distribution. Unlike the OU system, this occurs for very low values of $C$ even though the observable range is much larger. As a result, the intuitive choice of $C$ for thresholds in the tail of the distribution cannot be used. The values of $C$ chosen here are taken from preliminary work related to \cite{Bouchet_Wouters}.
 
 A related issue is the number of initial particles required to give an accurate return time curve. Relative error arguments for GPA do not hold here both because the optimal tilting value $C$ to threshold pair is nontrivial for complex systems and because the value $C$ cannot be chosen arbitrarily large. An alternative to this issue is to choose large enough initial particles $N$ so that relative error is only affected by the standard deviation of the tail probability estimates $\hat{\gamma}$ (see. sec. \ref{OU}); however, this number is nontrivial as convergence depends on how far the optimal value is from the chosen tilting value. 
 
GEV and GPA methods are able to estimate longer return times compared to MC brute force methods for the Lorenz 96 system. GEV has the advantage of maintaining the same relative error growth while difficulties in the optimal choice of $C$ and initial values cause probability tail estimates from GPA to have much larger relative error. Furthermore, GEV likelihood estimation requires a single run to estimate the optimal return level plot with confidence intervals where relative error can be approximated by the standard brute force growth rate ($\approx 1/\sqrt{N}\gamma_A$). On the other hand, GPA requires many runs to estimate the relative error and return level plot for threshold values that do not correspond to the center (or near center) of the $C$-shifted distribution.

\subsection{Planet Simulator (PlaSim)}

 We now describe a climate model on which our analysis will focus---\emph{Planet Simulator} ({\bf PlaSim}): a planet simulation model of intermediate complexity developed by the Universit\"{a}t Hamburg Meteorological Institute \cite{FKL}. Like most atmospheric models, PlaSim is a simplified model derived from the Navier Stokes equation in a rotating frame of reference. The model structure is given by five main equations which allow for the conservation of mass, momentum, and energy. For a full list of the variables used in the following equations please see table \ref{tab:PUMA}. The key equations are as follows:
\begin{itemize}
\item Vorticity Equation
\[
\frac{\partial\zeta}{\partial t} = \frac{1}{1-\mu^2}\frac{\partial}{\partial\lambda}\mathcal{F}_v-\frac{\partial}{\partial\mu}\mathcal{F}_u - \frac{\xi}{\tau_F}-K(-1)^h\bigtriangledown^{2h}\xi \tag{1}
\]
\item Divergence Equation
\[
\frac{\partial D}{\partial t} = \frac{1}{1-\mu^2}\frac{\partial}{\partial\lambda}\mathcal{F}_u+\frac{\partial}{\partial\mu}\mathcal{F}_v-\bigtriangledown^2\big(\frac{U^2+V^2}{2(1-\mu^2)}+\Phi+T_R\ln p_s\big)-\frac{D}{\tau_F}-K(-1)^h\bigtriangledown^{2h}D \tag{2}
\]
\item Thermodynamic Equation
\[
\frac{\partial T'}{\partial t} = -\frac{1}{(1-\mu^2)}\frac{\partial}{\partial\lambda}(UT')-\frac{\partial}{\partial\mu}(VT')+DT'-\dot{\sigma}\frac{\partial T}{\partial\sigma}+\kappa\frac{T\omega}{p}+\frac{T_R-T}{\tau_R}-K(-1)^h\bigtriangledown^{2h} T' \tag{3}
\]
\item Continuity Equation
\[
\frac{\partial(\ln p_s)}{\partial t} = -\frac{U}{1-\mu^2}\frac{\partial(\ln p_s)}{\partial\lambda}-V\frac{\partial(\ln p_s)}{\partial\mu}-D-\frac{\partial\dot{\sigma}}{\partial\sigma} \tag{4}
\]
\item  Hydrostatic Equation
\[
\frac{\partial\Phi}{\partial(\ln\sigma)} = -T \tag{5}
\]
\end{itemize}

Here,
\[
\begin{aligned}
U &= u\cos\phi-u\sqrt{1-\mu^2}, \quad V = v\cos\phi -v\sqrt{1-\mu^2},\\
\mathcal{F}_u &= V\zeta-\dot{\sigma}\frac{\partial U}{\partial\sigma}-T'\frac{\partial(ln p_s)}{\partial\lambda}, \quad
\mathcal{F}_v = -U\zeta-\dot{\sigma}\frac{\partial V}{\partial\sigma}-T'(1-\mu^2)\frac{\partial(\ln p_s)}{\partial\mu}.
\end{aligned}
\]

The combination of vorticity $(1)$ and divergence $(2)$ equations ensure the conservation of momentum in the system while the continuity equation $(4)$ ensures conservation of mass. The hydrostatic equation $(5)$ describes air pressure at any height in the atmosphere while the thermodynamic equation $(3)$ is essentially derived from the ideal gas law .

 \begin{table}
\caption{List of variables used in PUMA.\label{tab:PUMA}}
\centering
\begin{small}
\begin{tabular}{lcclcc|}
\hline
\hline
$\zeta$ & absolute vorticity & $\lambda$ & longitude\\
$\xi$ & relative vorticity & $\phi$ & latitude\\
$D$ & divergence & $\mu$ & $\sin(\phi)$\\
$\Phi$ & geopotential & $\kappa$ & adiabatic coefficient \\
$\omega$ & vertical velocity & $\tau_R$ & timescale of Newtonian cooling\\
$p$ & pressure & $\tau_F$ & timescale of Rayleigh friction\\
$p_s$ & surface pressure & $\sigma$ & vertical coordinate $p/p_s$\\
$K$ & hyperdiffusion & $\dot{\sigma}$ & vertical velocity $d\sigma/dt$\\
$u$ & zonal wind & $v$ & meridional wind\\
$h$ & hyperdiffusion order & $T_R$ & restoration temperature\\
$T$ & temperature & $T'$ & $T-T_R$\\
\hline
\end{tabular}
\end{small}
\end{table}

The  equations above are solved numerically with discretization given by a (variable) horizontal Gaussian grid~\cite{HS} and a vertical grid of equally spaced levels so that each grid-point has a corresponding latitude, longitude and depth triplet. (The default resolution is 32 latitude grid points, 64 longitude grid points and 5 levels.) At every fixed time step $t$ and each grid point, the atmospheric flow is determined by solving the set of model equations through the spectral transform method which results in a set of time series describing the system; including temperature, pressure, zonal, meridional and horizontal wind velocity, among others. The resulting time series can be converted through the PlaSim interface into a readily accessible data file (such as netcdf) where further analysis can be performed using a variety of platforms.  We refer to \cite{FKL} for more information. 


\subsubsection{GKLT, GEV and MC.}

Our observable of interest in PlaSim is the time series of summer European spatial average temperature anomalies. For simplicity, we set the climate boundary data to consistent July 1st conditions and remove the diurnal and annual cycles. This allows for perpetual summer conditions and saves on computational time. We define the European spatial average as the average over the set of 2-dimensional latitude and longitude pairs on the grid located between $36^\circ N-70^\circ N$ and $11^\circ W-25^\circ E$. Spatial average values are taken at 6 hour intervals. We subtract the long-run mean to obtain the sequence of summer European spatial average temperature anomalies used in this analysis.

We perform the GKLT algorithm on the European spatial averaged temperature time-series by considering initial values as the beginning of a year (360 days) to ensure each initial value is independent. It is important to note that initial values may be taken at much shorter intervals. We choose one year intervals because this initial data was readily available from the long control run. We estimate the resampling time $\tau = 8$ days as the approximate time for autocorrelation to reach near zero. For each experiment, we use 100 years (100 initial values) run for 17 complete steps of the GKLT algorithm, or 136 days, to estimate anomaly recurrence times for the $T = 8$-day time average. We remark that the choice of $T$ and $\tau$ here are the same, however this is not a requirement of the algorithm as illustrated in the Ornstein-Uhlenbeck system in sec. \ref{OU}. Results are compared to a 400 year (144,000 day) control run.  Added noise to ensure divergence of cloned trajectories is sampled uniformly from (preprogrammed noise) $[-\epsilon\sqrt{2},\epsilon\sqrt{2}]$ where $\epsilon = O(10^{-4})$.

Six experiments of the GKLT algorithm are performed on a starting ensemble of $N=100$ trajectories with initial values taken as the starting value of the European spatial average at the beginning of each year. The values $C = [0.01,0.05]$ (3 experiments per $C$ value) are chosen to tilt the distribution of the spatial-time average at resampling times $\tau=8$ days. We remark that constants $C=[0.1,2]$ are also tested with less favorable results; however, these tests were not included in the total numerical cost of MC brute force and GEV methods. We choose the observable described by eq. (\ref{obs2}), with $\phi(x_n(t))$ taken as the European spatial average temperature, to estimate return time curves of the 8-day time average of European spatial averaged temperature.

\begin{figure}
	\includegraphics[width=\textwidth]{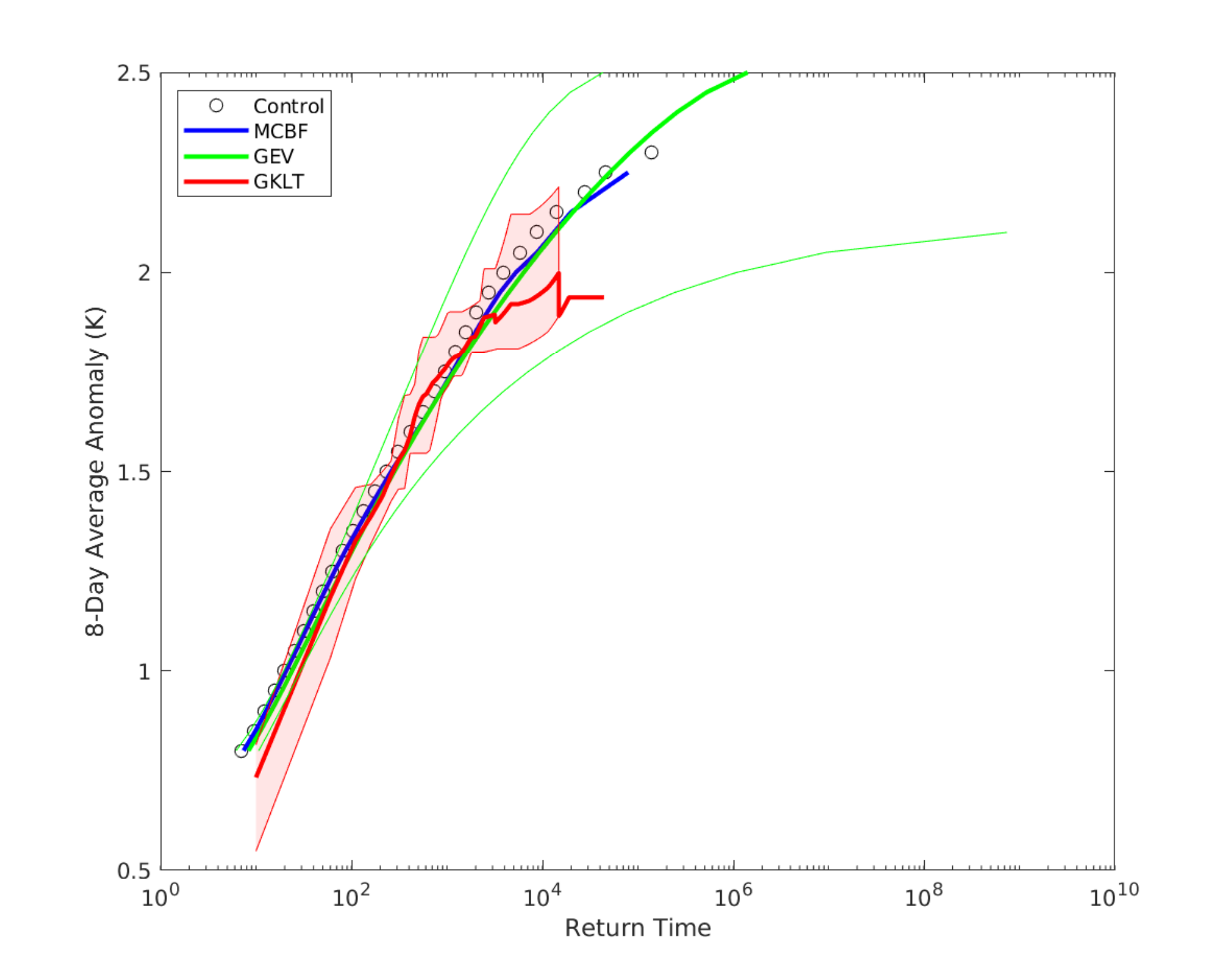}
	\caption{Return time estimates for 8-day average temperature anomalies from PlaSim using GKLT for $C=5\cdot 10^{-2}$ for $N=100$ over $136$ days starting particles, GEV and Monte Carlo estimates are provided with numerical cost $6\times100\times 136$ days. The control return time curve comes from a long brute-force run of $144,000$ days. Green outer lines indicate the 95\% confidence interval of the GEV. Red filled region indicates the deviation of the GKLT algorithm estimated over 6 runs.\label{fig.GKLT_PLASIM}}
\end{figure}

We refer to fig. \ref{fig.GKLT_PLASIM} for this discussion. GEV and MC methods agree almost completely up to return times of $1\cdot 10^6$ with the GEV continuing to provide estimates for longer return times. 95\% confidence intervals for the GEV (green thin lines) are a result of the likelihood function. The return time curve for GKLT is formed by the set of return time values from each of the 6 experiments that fall within 1/2 standard deviation of the mean of the shifted distribution. Hence, the deviation for GKLT (red region) is estimated by the minimum and maximum deviation of anywhere between 2 and 6 return time values for each threshold. Compared to that of the long control run, GKLT provides reliable estimates for return times up to $1\cdot 10^4$, while GEV estimates remain near those of the long control run for return times up to $1\cdot 10^6$. Deviation estimates for GKLT are smaller than the 95\% confidence interval for the GEV for return times longer than $1\cdot 10^3$; however, this may be the result of a low number of experiments. We also remark that the deviation estimate of the GKLT method for return times of the 8-day average anomaly near 1.5 Kelvin are much smaller compared to other thresholds. This reduction suggests that at least one of the $C$ values chosen in GKLT is close to optimal for the 1.5 Kelvin threshold.

\section{Discussion}

In this paper we have discussed two importance sampling (IS) methods: Genealogical particle analysis (GPA) which is used to estimate tail probabilities of a time series of observations at a fixed sampling rate, and GKLT which is used to estimate tail probabilities of a corresponding time average. Both methods work by tilting the distribution of observations in a reversible way so that the rare events corresponding to tail probabilities are sampled more often. We have illustrated the particular case when the observations of interest are distributed according to a symmetric, heavy-tailed distribution and a rare event consists of an exceedance of a certain level where the natural choice of tilt corresponds to a shift towards the tail.

We compare results of these two methods with classical statistics where rare event estimation is given by the Generalized Extreme Value (GEV) distribution. Under the goal of obtaining a return level curve, we have shown that the GEV outperforms both IS methods for all three systems used in this analysis by providing generally lower relative error and longer return time estimates. We have also illustrated a few disadvantages in IS methods including the strict dependence of the tilting value to initial conditions and requirement of multiple runs for return time curve and relative error estimation while demonstrating that classical GEV results only require a single run to estimate return time curves and follow standard brute force relative error growth. On the other hand, we have shown that our results do not conflict with previous literature and that both the GEV and IS methods outperform Monte Carlo brute force methods in estimating longer return times. In fact, following previous literature we have shown that IS methods can result in lower relative error than that of the GEV on subsets of tail probabilities (and hence, that of MC brute force) provided the optimal tilting value can be chosen.

In general, these results support the idea of using GEV methods over IS under the condition that optimal tilting values cannot be determined a priori and/or return time curves, rather than returns for a single level, are of interest. We emphasize that these results should not be taken to discount the value of importance sampling. The power of these methods can be seen in the decrease in relative error when optimal tilting values can be chosen. It would be interesting to see more theoretical work in estimating such values which, at the moment, requires an explicit formula of the (unknown) distribution of the observable. Other numerical work can also be completed using IS methods which does not involve tail probability estimation. One particular perspective we plan to explore is the algorithms' ability to provide the set of trajectories which most likely end in an extreme event.

\clearpage

\paragraph{Acknowledgements}
We warmly thank Frank Lunkeit at  Universit{\"a}t  Hamburg for very helpful discussions and advice concerning PlaSim.
 MN was supported in part by NSF Grant DMS 1600780.

\end{document}